\documentclass[reprint,amsmath,amssymb,showpacs,aps]{revtex4-1}

\usepackage{graphicx}

\newcommand*{\ie}{i.\,e.}
\newcommand*{\eg}{e.\,g.}

\newcommand*{\bs}{\boldsymbol}
\newcommand*{\mbb}{\mathbb}

\newcommand*{\diff}{\mathop{}\!\mathrm{d}}
\newcommand*{\pd}{\partial}
\newcommand*{\im}{\ensuremath{\mathrm{i}}}
\newcommand*{\e}{\mathop{\mathrm{e}}\nolimits}
\newcommand*{\const}{\ensuremath{\mathrm{const}}}
\DeclareMathOperator{\grad}{\nabla\!}
\DeclareMathOperator{\diverg}{div\!}
\newcommand*{\sphere}{\ensuremath{{\mbb{S}^2}}}
\newcommand*{\average}[1]{\left<#1\right>}

\newcommand*{\Four}{\mathcal{F}}
\newcommand*{\de}{\mathrm{DE}}
\newcommand*{\te}{\mathrm{TE}}
\newcommand*{\jeje}{\mathrm{Je}}
\newcommand*{\jetp}{\mathrm{TP}}

\newcommand*{\boltz}{\mathrm{Bo}}
\newcommand*{\regul}{\mathrm{r}}
\newcommand*{\singul}{\mathrm{s}}
\newcommand*{\steady}{\infty}

\numberwithin{equation}{section}

\begin{document}

\preprint{APS/123-QED}

\title{Local immobilization of particles in mass transfer\\ described by the equation of the Jeffreys type}

\author{S.\,A.\,Rukolaine}%
\email{rukol@ammp.ioffe.ru}
\author{A.\,M.\,Samsonov}%
\email{samsonov@math.ioffe.ru}
\affiliation{%
  The Ioffe Physical Technical Institute of the Russian Academy of Sciences, 26~Polytekhnicheskaya, St.\,Petersburg, 194021, Russia
}%

\date{\today}

\begin{abstract}
  We consider the equation of the Jeffreys type as the basic one in three different models of mass transfer, namely, the \emph{Jeffreys type} and \emph{two-phase} models, and the \emph{$D_1^{}$ approximation to the linear Boltzmann equation}. We study two classic $1+1$D problems in the framework of each model. The first problem is the transfer of a substance initially confined in a point. The second problem is the transfer of a substance from a stationary point source. We calculate the mean-square displacement (MSD) for the solutions of the first problem. The temporal behaviour of the MSD in the framework of the first and third models is found to be the same as that in the Brownian motion described by the standard Langevin equation. Besides, we find a remarkable phenomenon when a portion of the substance \emph{does not move}.
\end{abstract}


\maketitle

\section{Introduction}

The classic diffusion equation, based on Fick's law, is  widely used
for approximate description of non-anomalous diffusion (dispersion
of a substance or species)~\cite{KondepudiPrigogine:1998,
BirdEtAl:2002} and Brownian motion~\cite{Risken:1989, Mazo:2002,
CoffeyEtAl:2004}. However, Fick's law neglects the mass (inertia) of
moving particles (molecules), and, therefore, the diffusion equation
gives an appropriate and accurate model for diffusion phenomena only
in weakly inhomogeneous media and/or for processes, which are slow
in time, when relaxation time is short compared to a characteristic
time scale. Otherwise the description of diffusion by the diffusion
equation may fail \cite{Rice:1985}. Many biological media, \eg,
cellular cytoplasm, are strongly inhomogeneous,
therefore, diffusion in them is not Fickian and its description by the diffusion equation is questionable.

Note that the counterpart of Fick's law
is Fourier's law in the theory of heat conduction~\cite{Fourier:2009,
KondepudiPrigogine:1998, BirdEtAl:2002}. The latter leads to the heat
equation, similar to the diffusion equation. Fick's law was postulated
by analogy with Fourier's law, which was proposed first~\cite{BirdEtAl:2002}.

The simplest modification of Fick's law, taking into account the
inertia  of moving particles, is \emph{Cattaneo's equation}
\cite{JosephPreziosi:1989, JosephPreziosi:1990, Straughan:2011}.
Strictly speaking, Cattaneo's equation modifies Fourier's law, since
Cattaneo considered heat conduction, however, one can apply this to
mass transfer as well. The modification
leads to the \emph{telegraph equation}, providing the finite speed
of propagation \cite{MorseFeshbach:1953v1, MoninYaglom:1971, JosephPreziosi:1989,
JosephPreziosi:1990, Weiss:2002, JouEtAl:2010}. The telegraph
equation was proposed to be a substitution of the diffusion and heat
equations. However, both two- and three-dimensional
telegraph equations meet formal obstacles
since solutions to initial value problem for them
may become negative \cite{PorraEtAl:1997, KornerBergmann:1998}.

Long before Cattaneo H.\,Jeffreys proposed a relation for
rheological description of Earth's core \cite{Jeffreys:1929,
Reiner:1958}, that can be considered as a combination of
Fick's (or Fourier's) law and Cattaneo's equation. We define this
the \emph{law (relation) of the Jeffreys type}. This law leads to a
partial differential equation of the \emph{third} order, called the
\emph{equation of the Jeffreys type} \cite{JosephPreziosi:1989,
JosephPreziosi:1990}, also known as the simplest equation of the
dual phase lag model of heat conduction \cite{Tzou:1997,
TzouXu:2011, Straughan:2011}.
We call this the \emph{Jeffreys type} model. This model was used for description of viscoelastic fluids \cite{PreziosiJoseph:1987,
JosephPreziosi:1989, JouEtAl:2011}, Taylor dispersion \cite{Camacho:1993, JouEtAl:2011} and heat transfer \cite{JosephPreziosi:1989,
JosephPreziosi:1990, Tzou:1997, Sobolev:1997, Zhang:2009, JouEtAl:2010,
LiuChen:2010, TzouXu:2011, Straughan:2011, LiuEtAl:2012,
VanFulop:2012}.

There is another \emph{two-temperature} \cite{Tzou:1997, Sobolev:1997, JouEtAl:2010,
TzouXu:2011, Straughan:2011} or \emph{two-phase} \cite{Crank:1975, Beaudouin:2006, CoppeyEtAl:2007, Bancaud:2009} model, which leads also to the equation of the Jeffreys type.
This model is qualitatively different from the Jeffreys type one.
Nevertheless, to the best of our knowledge, there is no clear distinction between cases when the equation of the Jeffreys type describes the Jeffreys type and the two-temperature (two-phase) models \cite{Sobolev:1997, JouEtAl:2010, TzouXu:2011}.
Moreover, in Ref.\,\cite{Sobolev:1997} the behaviour of the two-temperature model is illustrated by that of the Jeffreys type model.
And in Ref.\,\cite{WangWei:2008} the authors erroneously state that the two models are equivalent.

The diffusion equation is known to be the simplest approximation to the linear Boltzmann equation~\cite{DuderstadtMartin:1979}, and the latter can be considered as a mesoscopic model of the former.
It is notable that the telegraph equation is the $P_1$ approximation (the next after the diffusion one) to the linear Boltzmann equation~\cite{DuderstadtMartin:1979}.
Recently, $D_N^{}$ approximations to the linear Boltzmann equation were proposed~\cite{SchaferEtAl:2011}. They generalize the classic diffusion approximation, which corresponds to $N=0$. We have found that the \emph{$D_{N=1}^{}$ approximation} (the next after the telegraph one) leads to the equation of the Jeffreys type, the model being qualitatively different from both the Jeffreys type and two-phase ones. Note that this model is similar to the one of Guyer and Krumhansl in the theory of second sound
\cite{GuyerKrumhansl:1966a, GuyerKrumhansl:1966b, Straughan:2011,
JouEtAl:2010}.

The primary motivation of this study was to investigate the equation of the Jeffreys type as a substitution of the diffusion equation instead of the telegraph one for description of mass transfer.
Eventually in this paper we study and compare the three models (the \emph{Jeffreys type} and \emph{two-phase} models, and the \emph{$D_{N=1}^{}$ approximation to the linear Boltzmann equation}) as models of mass transfer.
We study two classic $1+1$D problems, typical for mass transfer, in the framework of each model, where, as far as we know, the problems have not been studied.
The first problem is the transfer of a substance
initially confined in a point. The second problem is the transfer
of a substance from a stationary point source. We calculate the
mean-square displacement (MSD) for the solutions of the first
problem, because the MSD is an integral quantity whose temporal dependence characterizes diffusion and Brownian motion. The temporal behaviour of the MSD in the framework of the
first and third models is found to be the same as that in the
Brownian motion described by the standard Langevin equation. We
remind here that the behaviour of the MSD in the framework of the
diffusion equation is wrong at small values of time, where it must be
ballistic. Besides, we find a remarkable phenomenon when a portion
of the substance \emph{does not move}.

The rest of the paper is organized as follows.
In Section~\ref{sec:DiffEq} we briefly remind the phenomenological derivation of the diffusion equation.
In Section~\ref{sec:TeleEq}
we briefly recall some facts about the telegraph equation.
In Section~\ref{sec:JeffEq} we describe the models of mass transfer
related to the equation of the Jeffreys type. In
Section~\ref{sec:IVPJeffEqNoSource} we study the diffusion of a
substance initially confined in a point in the framework of the
three models. In Section~\ref{sec:MSD} we calculate the mean-square
displacement for the solutions of the problems, considered in
Section~\ref{sec:IVPJeffEqNoSource}. In
Section~\ref{sec:IVPJeffEqSource} we study the diffusion of a
substance from a stationary point source also in the framework of
the three models. Section~\ref{sec:Concl} contains some concluding remarks.

\section{Diffusion equation}
\label{sec:DiffEq}

The macroscopic law of mass balance for a substance is expressed by the continuity equation \cite{Murray:2002v1}
\begin{equation}
  \label{eq:ContinEqSource}
  \frac{\pd u}{\pd t} + \diverg \bs{J}
  = f,
\end{equation}
where $u \equiv u(\bs{x},t)$ is the concentration of the substance, $\bs{x} = (x_1,x_2,x_3)$ is a point, $t$ is time, $\bs{J} \equiv \bs{J}(\bs{x},t)$ is the flux of the substance, $f \equiv f(\bs{x},t,u)$ is the net rate of production or absorption (degradation) of the substance.

In the simplest approximation the flux is related to
the concentration by phenomenological Fick's (first)
law~\cite{KondepudiPrigogine:1998, BirdEtAl:2002, Murray:2002v1}
\begin{equation}
  \label{eq:FickLaw}
  \bs{J} = - D \grad u,
\end{equation}
where $D$ is the diffusion coefficient.

The continuity equation~\eqref{eq:ContinEqSource} and Fick's law
lead to the reaction diffusion equation
\begin{equation}
  \label{eq:DiffEqSource}
  \frac{\pd u}{\pd t} - D \Delta u =
  f.
\end{equation}
To determine a unique solution of the diffusion equation one imposes the initial condition
\begin{equation}
  \label{eq:InitCondDiffEq}
  \left. u \right|_{t=0}^{} = u_0^{},
\end{equation}
where $u_0^{} \equiv u_0^{}(\bs{x})$ is the distribution of the concentration at time $t=0$.

Note that the diffusion equation is the simplest approximation to
the linear Boltzmann equation~\cite{DuderstadtMartin:1979}, see
Eq.\,\eqref{eq:BoltzEqDiffApprox}.

\section{Telegraph equation}
\label{sec:TeleEq}

Fick's law neglects the inertia of moving particles.
Cattaneo's equation \cite{JosephPreziosi:1989, JosephPreziosi:1990, Straughan:2011}
\begin{equation}
  \label{eq:CattEq}
  \tau \frac{\pd \bs{J}}{\pd t} + \bs{J}
  =
  - D \grad u,
\end{equation}
where $\tau$ is the relaxation time, modifies Fick's law, taking the inertia into account.
Indeed, Cattaneo's equation can be written in the equivalent integral form
\begin{equation}
  \label{eq:CattEqInt}
  \bs{J} =
  - \frac{D}{\tau} \int_0^t \e^{-(t - t')/\tau} \grad u(\bs{x},t') \diff t' + \e^{-t/\tau} \bs{J}_0^{},
\end{equation}
where $\bs{J}_0^{} \equiv \bs{J}_0^{}(\bs{x})$ is the distribution
of flux at time $t=0$. Eq.\,\eqref{eq:CattEqInt} shows that Cattaneo's equation takes into account the
prehistory of a process, since flux depends on the gradient of the
concentration at earlier time, the dependence being exponentially
decreasing with time.
If the relaxation time $\tau$ in Cattaneo's
equation tends to zero, one obtains in the limit Fick's law.

The continuity equation \eqref{eq:ContinEqSource} and Eq.\,\eqref{eq:CattEqInt} lead to the integro-differential equation
\begin{equation*}
  \frac{\pd u}{\pd t} -  \frac{D}{\tau} \int_0^t \e^{-(t - t')/\tau} \Delta u(\bs{x},t') \diff t' + \e^{-t/\tau} \diverg \bs{J}_0^{} =
  f.
\end{equation*}
This equation with the initial condition~\eqref{eq:InitCondDiffEq} is equivalent to the reaction telegraph (or damped wave) equation~\cite{Hadeler:1999, Mendez:2010}
\begin{equation}
  \label{eq:TeleEqSource}
  \tau \frac{\pd^2 u}{\pd t^2} +
  \left( 1 - \tau \frac{\pd f}{\pd u} \right) \frac{\pd u}{\pd t} -
  D \Delta u =
  f +
  \tau \frac{\pd f}{\pd t},
\end{equation}
with the initial conditions
\begin{equation}
  \label{eq:InitCondTeleEqSource}
  \left. u \right|_{t=0}^{} = u_0^{},
  \quad
  \left. \frac{\pd u}{\pd t} \right|_{t=0}^{} =
  - \diverg \bs{J}_0^{} + f_0^{},
\end{equation}
where $f_0^{} \equiv f_0^{}(\bs{x},u_0^{}) = \left. f \right|_{t=0}^{}$ is the distribution of sources at time $t=0$.
If $\tau =0$, the telegraph equation~\eqref{eq:TeleEqSource} becomes the diffusion equation~\eqref{eq:DiffEqSource}.

The telegraph equation can also be obtained as the $P_1$
approximation to the linear Boltzmann equation~\cite{DuderstadtMartin:1979}, see
Eq.\,\eqref{eq:BoltzEqTeleApprox}.

The telegraph equation is hyperbolic, providing the finite speed of
signal propagation, and it was proposed to be a substitution of the
parabolic diffusion and heat equations \cite{MoninYaglom:1971,
JosephPreziosi:1989, JosephPreziosi:1990, Weiss:2002, JouEtAl:2010}.
However, two- and three-dimensional telegraph equations have a
formal flaw since their solutions may take negative values even if
the initial values are positive \cite{PorraEtAl:1997,
KornerBergmann:1998}. Besides, the applicability of the telegraph
equation to the description of heat transfer is
doubtful~\cite{KornerBergmann:1998, BrightZhang:2009, ZhangEtAl:2011}.

\section{Equation of the Jeffreys type}
\label{sec:JeffEq}

\subsection{Jeffreys type model}

The relation, combining Fick's law and Cattaneo's equation, has the form \cite{JosephPreziosi:1989, JosephPreziosi:1990}
\begin{multline}
  \label{eq:JeffLaw}
  \tau \frac{\pd \bs{J}}{\pd t} + \bs{J} =
  - \tau D_1^{} \frac{\pd \grad u}{\pd t} - \left( D_1^{} + D_2^{} \right) \grad u\\
  \equiv
  - \left( D_1^{} + D_2^{} \right) \left[ \tau_2^{} \frac{\pd \grad u}{\pd t} + \grad u \right],
\end{multline}
where $D_1^{} > 0$, $D_1^{} + D_2^{} > 0$ and
\begin{equation*}
  \tau_2^{} =
  \frac{\tau D_1^{}}{D_1^{} + D_2^{}}
\end{equation*}
is another relaxation time, or, equivalently, the integro-differential form
\begin{multline}
  \label{eq:JeffLawInt}
  \bs{J} =
  - D_1^{} \grad u - \frac{D_2^{}}{\tau} \int_0^t \e^{-(t - t')/\tau} \grad u(\bs{x},t') \diff t'\\
  + \e^{-t/\tau} \left( D_1^{} \grad u_0^{} + \bs{J}_0^{} \right),
\end{multline}
where $u_0^{} \equiv u_0^{}(\bs{x})$ and $\bs{J}_0^{} \equiv
\bs{J}_0^{}(\bs{x})$ are the distributions of the concentration and
flux, respectively, at time $t=0$.
We name the
relations~\eqref{eq:JeffLaw} and \eqref{eq:JeffLawInt} the law of
the Jeffreys type after H.\,Jeffreys who proposed similar relations
for rheological description of the Earth core \cite{Jeffreys:1929,
Reiner:1958}. Fick's law and Cattaneo's equation are particular
cases of the law of the Jeffreys type. Indeed, if $\tau$ in
Eqs.\,\eqref{eq:JeffLaw} and \eqref{eq:JeffLawInt} tends to zero,
one obtains in the limit Fick's law with $D = D_1^{} + D_2^{}$,
while $D_1^{} = 0$ leads to Cattaneo's equation.

The law of the Jeffreys type \eqref{eq:JeffLaw} includes two
different cases, $\tau > \tau_2^{}$ and $\tau < \tau_2^{}$,
depending on whether the relaxation time $\tau$ is higher or lower
than $\tau_2^{}$. Both cases are considered in literature, see, \eg,
\cite{Tzou:1997, Zhang:2009, LiuChen:2010, LiuEtAl:2012}. The first
inequality $\tau
> \tau_2^{}$ is equivalent to $D_2^{} > 0$. In this case the
relation \eqref{eq:JeffLawInt} means that \emph{flux is determined
by the the concentration gradient} both at the same moment and
preceding time, the dependence on the past being exponentially
decreased. The second inequality $\tau < \tau_2^{}$ is equivalent to
$D_2^{} < 0$. In this case the law of the Jeffreys type can be
written in the equivalent form
\begin{multline}
  \label{eq:JeffLawInt2}
  \grad u =
  \frac{1}{D_1^{}} \bigg[ - \bs{J} + \frac{D_2^{}}{\tau D_1^{}} \int_0^t \e^{-(t - t')/\tau_2^{}} \bs{J}(\bs{x},t') \diff t'\\
  + \e^{-t/\tau_2^{}} \left( D_1^{} \grad u_0^{} +  \bs{J}_0^{} \right) \bigg],
\end{multline}
which means that the \emph{gradient of the concentration is
determined by flux} both at the same moment and preceding time, the
dependence on the past being exponentially decreased. Note that the
relation \eqref{eq:JeffLawInt2} can be used for setting boundary
conditions for $u$ if mass transfer is considered in a finite
domain.

The continuity equation~\eqref{eq:ContinEqSource} and the integro-differential law of the Jeffreys type~\eqref{eq:JeffLawInt} lead to the integro-differential equation
\begin{multline*}
  \frac{\pd u}{\pd t} - D_1^{} \Delta u - \frac{D_2^{}}{\tau} \int_0^t \e^{-(t - t')/\tau} \Delta u(\bs{x},t') \diff t'\\
  + \e^{-t/\tau} \left( D_1^{} \Delta u_0^{} + \diverg \bs{J}_0^{} \right) =
  f.
\end{multline*}
This equation with the initial condition~\eqref{eq:InitCondDiffEq}
is equivalent to the equation of the \emph{third} order
\begin{multline}
  \label{eq:JeffEqJeffModelSource}
  \tau \frac{\pd^2 u}{\pd t^2} +
  \left( 1 - \tau \frac{\pd f}{\pd u} \right) \frac{\pd u}{\pd t} -
  \tau D_1^{} \frac{\pd \Delta u}{\pd t} -
  \left( D_1^{} + D_2^{} \right) \Delta u\\
  =
  f +
  \tau \frac{\pd f}{\pd t},
\end{multline}
with the initial conditions~\eqref{eq:InitCondTeleEqSource}.
We name Eq.\,\eqref{eq:JeffEqJeffModelSource} the \emph{equation of the Jeffreys
type}~\cite{JosephPreziosi:1989}. The diffusion
equation~\eqref{eq:DiffEqSource} and the telegraph
equation~\eqref{eq:TeleEqSource} are the particular cases of the
equation of the Jeffreys type for $\tau =0$ and $D_1^{} = 0$,
respectively.

Eq.\,\eqref{eq:JeffLaw} can also be derived formally in the framework of the dual phase lag model~\cite{Tzou:1997, TzouXu:2011, Straughan:2011}. The model applies heat transfer and Fourier's law, however, one can extend this to mass transfer and Fick's law as well. In this framework Fick's law is replaced by the relation
\begin{equation}
  \label{eq:DualPhaseLag}
  \bs{J}(\bs{x}, t + \tau)
  =  - D \grad u (\bs{x}, t + \tau_2^{}),
\end{equation}
where $\tau$ and $\tau_2^{}$ are the time lags of the flux and the gradient of the concentration, respectively.
Both sides of the relation are expanded with the use of Taylor's formula. If only terms up to the first order are retained one obtains the relation
\begin{equation*}
  \tau \frac{\pd \bs{J}}{\pd t} + \bs{J}
  =
  - D \left( \tau_2^{} \frac{\pd \grad u}{\pd t} + \grad u \right),
\end{equation*}
which is nothing but Eq.\,\eqref{eq:JeffLaw} with $D_1^{} + D_2^{} = D$.
Note that if $\tau_2^{} = 0$ one obtains single phase lag model and Cattaneo's equation.

However, Eq.\,\eqref{eq:DualPhaseLag} (both for $\tau_2^{} > 0$ and $\tau_2^{} = 0$) together with the continuity equation yields delay equations leading to ill-posed initial value problems (with unstable solutions) \cite{JordanEtAl:2008, DreherEtAl:2009}.
Therefore, the phase lag models cannot be considered as sensible physical ones. At the same time the formal ``approximations'' to the phase lag models lead to well-posed initial value problems.

\subsection{Two-phase (two-temperature) model}

In this model two phases of a substance (or species) are considered: free (mobile) and bound (immobile), see, \eg,~\cite{Crank:1975, Beaudouin:2006, CoppeyEtAl:2007, Bancaud:2009}. The concentrations of these substances are denoted by $u \equiv u(\bs{x},t)$ and $v \equiv v(\bs{x},t)$, respectively, and satisfy the reaction diffusion system
\begin{subequations}
  \label{eq:TwoPhModelSource}
  \begin{alignat}{2}
    & \dfrac{\pd u}{\pd t} - D_1^{} \Delta u + k_1^{} u - k_2^{} v && = f,\label{eq:TwoPhModelSourceEqU}\\[1ex]
    & \dfrac{\pd v}{\pd t} + k_2^{} v - k_1^{} u && = 0,\label{eq:TwoPhModelEqV}
  \end{alignat}
\end{subequations}
with the initial conditions
\begin{equation}
  \label{eq:InitCondTwoPhModel}
  \left. u \right|_{t=0}^{} = u_0^{},
  \quad
  \left. v \right|_{t=0}^{} = v_0^{},
\end{equation}
where $D_1^{}$ is the diffusion coefficient of the free substance,
$k_1$ and $k_2$ are the coefficients of interphase mass transfer, $f
= f(\bs{x},t,u)$ is the net rate of production and
absorption (degradation) of the free substance, $v_0^{} \equiv
v_0^{}(\bs{x})$ is the the distribution of the concentration of the
immobile substance at time $t = 0$. The coefficients $k_1$ and $k_2$
are evidently \emph{positive} in this model.

The concentration $v$ can be expressed through $u$ from the equation~\eqref{eq:TwoPhModelEqV} by
\begin{equation*}
  v =
  k_1^{} \int_0^t \e^{-k_2^{} (t - t')} u(\bs{x},t') \diff t' + \e^{-k_2^{} t} v_0^{}.
\end{equation*}
Then the equation~\eqref{eq:TwoPhModelSourceEqU} leads to the equation
\begin{multline*}
  \dfrac{\pd u}{\pd t} - D_1^{} \Delta u + k_1^{} u - k_1^{} k_2^{} \int_0^t \e^{-k_2^{} (t - t')} u(\bs{x},t') \diff t'\\
  - k_2^{} \e^{-k_2^{} t} v_0^{} =
  f.
\end{multline*}
This equation with the first of the conditions~\eqref{eq:InitCondTwoPhModel} is equivalent to the \emph{equation of the Jeffreys type}
\begin{multline}
  \label{eq:JeffEqTwoPhModelSourceU}
  \frac{\pd^2 u}{\pd t^2} +
  \left( k_1^{} + k_2^{} - \frac{\pd f}{\pd u} \right) \frac{\pd u}{\pd t} -
  D_1^{} \frac{\pd \Delta u}{\pd t} -
  k_2^{} D_1^{} \Delta u\\
  =
  k_2^{} f +
  \frac{\pd f}{\pd t},
\end{multline}
with the initial conditions
\begin{equation}
  \label{eq:InitCondJeffEqTwoPhModelSourceU}
  \left. u \right|_{t=0}^{} =
  u_0^{},
  \quad
  \left. \frac{\pd u}{\pd t} \right|_{t=0}^{} =
  D_1^{} \Delta u_0^{} - k_1^{} u_0^{} + k_2^{} v_0^{}  + f_0^{}.
\end{equation}

The equation for $v$ is
\begin{multline}
  \label{eq:JeffEqTwoPhModelSourceV}
  \frac{\pd^2 v}{\pd t^2} + \left( k_1^{} + k_2^{} \right) \frac{\pd v}{\pd t} - D_1^{} \frac{\pd \Delta v}{\pd t} - k_2^{} D_1^{} \Delta v\\
  =
  k_1^{} f \!\left( \!\bs{x}, t, \frac{1}{k_1^{}} \!\left( \frac{\pd v}{\pd t} +k_2^{} v \right) \!\right),
\end{multline}
which is different from the equation
\eqref{eq:JeffEqTwoPhModelSourceU}, if $f \neq 0$.
The initial conditions for $v$ are
\begin{equation}
  \label{eq:InitCondJeffEqTwoPhModelSourceV}
  \left. v \right|_{t=0}^{} =
  v_0^{},
  \quad
  \left. \frac{\pd v}{\pd t} \right|_{t=0}^{} =
  k_1^{} u_0^{} - k_2^{} v_0^{}.
\end{equation}

The counterpart of the two-phase model in the field of heat transfer is the two-temperature model \cite{KaganovEtAl:1957, Sobolev:1997, Tzou:1997, BrightZhang:2009}.

\subsection{Relations between the coefficients of the two models}

If sources and sinks (absorption) are absent, \ie, $f = 0$, then the
equations of the Jeffreys type \eqref{eq:JeffEqJeffModelSource} and
\eqref{eq:JeffEqTwoPhModelSourceU} are identical, and the
coefficients are related by
\begin{subequations}
  \label{eq:RelCoeffJeffTwoPhModels}
  \begin{gather}
    \tau =
    \frac{1}{k_1^{} + k_2^{}}
    \quad\text{and}\quad
    D_2^{} =
    - \frac{k_1^{}}{k_1^{} + k_2^{}} D_1^{},
    \label{eq:RelCoeffJeffTwoPhModelsA}\\
    \intertext{or, vice versa,}
    k_1^{} =
    - \frac{1}{\tau} \frac{D_2^{}}{D_1^{}} \equiv
    \frac{1}{\tau} - \frac{1}{\tau_2^{}}\hspace{11em}\notag\\
    \hspace{6em}\text{and}\quad
    k_2^{} =
    \frac{1}{\tau} \left( 1 + \frac{D_2^{}}{D_1^{}} \right) \equiv
    \frac{1}{\tau_2^{}},
    \label{eq:RelCoeffJeffTwoPhModelsB}
  \end{gather}
\end{subequations}
the diffusion coefficient $D_1^{}$ being the same in the two models.
At the same time, the \emph{initial conditions}~\eqref{eq:InitCondTeleEqSource} and \eqref{eq:InitCondJeffEqTwoPhModelSourceU} for the equations, concerning the time derivative, are \emph{different}. Below, in Section~\ref{sec:IVPJeffEqNoSource}, it will be shown that this leads to qualitatively different behaviour of the solutions to the initial value problems for the equations of the Jeffreys type.

It is necessary to emphasize here that the \emph{positive}
coefficient $k_1^{}$ in the two-phase
model~\eqref{eq:TwoPhModelSource} corresponds to the
\emph{negative}(!) coefficent $D_2^{}$ in the law of the Jeffreys
type \eqref{eq:JeffLaw}, \eqref{eq:JeffLawInt},
\eqref{eq:JeffLawInt2}. Conversely, the \emph{positive} coefficient
$D_2^{}$ corresponds to the \emph{negative}(!) coefficient $k_1^{}$.

\subsection{$D_{N=1}^{}$ approximation to the linear Boltzmann
equation}

We consider here an approximation to the linear Boltzmann equation
(also referred to as the linear transport or radiative transfer
equation) \cite{DuderstadtMartin:1979, Cercignani:1988,
Modest:2003}, which describes, \eg, neutron transport and radiative
heat transfer (transport  of thermal energy by photons), see
Appendix~\ref{sec:BoltzEqD1Approx}. We use the notation $D_{N=1}^{}$
instead of $D_1^{}$, since the latter stands for the coefficient.

Consider the monoenergetic (one-speed) linear Boltzmann equation
\begin{multline}
  \label{eq:BoltzEq}
  \frac{\pd \psi}{\pd t} + c \,\bs\varOmega \cdot \grad \psi + \left( \kappa + \sigma \right) \psi\\
  =
  \sigma \!\int_\sphere K(\bs\varOmega \cdot \bs\varOmega') \psi(\bs{x},\bs\varOmega',t) \diff\bs\varOmega' + \frac{1}{4\pi} F,
\end{multline}
where $\psi \equiv \psi(\bs{x},\bs\varOmega,t)$ is the particle
phase space density, \ie, the density of particles at the point
$\bs{x}$ and at time $t$ moving along the direction $\bs\varOmega
\in \sphere$, $\sphere$ is the unit sphere in $\mbb{R}^3$, $c$ is
the velocity of particles, $\kappa$ and $\sigma$ are the absorption
and scattering rates, respectively, $K$ is the collision (or
scattering) kernel,
$F \equiv F(\bs{x},t)$ is the source density (due to isotropic sources for simplicity).

Integration of the linear Boltzmann equation over the unit sphere, together with the normalization $\int_\sphere K(\bs\varOmega \cdot \bs\varOmega') \diff\bs\varOmega = 1$, gives the continuity equation
\begin{equation}
  \label{eq:BoltzEqMomentEqZero}
  \frac{\pd u}{\pd t} + \diverg \bs{J} + \kappa u =
  F,
\end{equation}
where
\begin{equation}
  \label{eq:BoltzDens}
  u(\bs{x},t) =
  \int_\sphere \psi(\bs{x},\bs\varOmega,t) \diff\bs\varOmega
\end{equation}
is the particle density,
and
\begin{equation}
  \label{eq:BoltzFlux}
  \bs{J}(\bs{x},t) =
  c \int_\sphere \bs\varOmega \,\psi(\bs{x},\bs\varOmega,t) \diff\bs\varOmega
\end{equation}
is flux. In the $D_{N=1}^{}$ approximation \cite{SchaferEtAl:2011}
the particle density and flux are related by the
equation~\eqref{eq:BoltzEqMomentEqTwo}, which can be written as
\begin{multline}
  \label{eq:BoltzEqMomentEqTwoNoAbsorpSourceTauD1D2}
  \tau \frac{\pd \bs{J}}{\pd t}
  + \bs{J}\\
  =
  - \left( D_1^{} + D_2^{} \right) \grad u
  + \frac{\tau D_1^{}}{4} \left( 3 \Delta \bs{J} + \grad \diverg \bs{J} \right),
\end{multline}
where
\begin{multline*}
  \tau
  = \frac{1}{\kappa + \sigma_1^{}},
  \quad
  D_1^{}
  = \frac{4 c^2}{15 (\kappa + \sigma_2^{})},\\
  D_2^{}
  = \left[ \frac{1}{\kappa + \sigma_1^{}} - \frac{4}{5 (\kappa + \sigma_2^{})} \right] \frac{c^2}{3}
  \quad\text{and}\quad
  \gamma = \kappa,
\end{multline*}
see Appendix~\ref{sec:BoltzEqApprox} (the parameter $\gamma$ will be
used elsewhere).

The continuity equation \eqref{eq:BoltzEqMomentEqZero} and relation
\eqref{eq:BoltzEqMomentEqTwoNoAbsorpSourceTauD1D2} imply that the
particle density satisfies the \emph{equation of the Jeffreys type}
\begin{multline}
  \label{eq:JeffEqBoltzModelAbsorpSource}
  \tau \frac{\pd^2 u}{\pd t^2}
  + \left( 1 + \tau \gamma \right) \frac{\pd u}{\pd t}
  - \tau D_1^{} \frac{\pd \Delta u}{\pd t}\\
  - \left[ \left( 1 + \tau \gamma \right) D_1^{} + D_2^{} \right] \Delta u
  + \gamma u\\
  =
  F
  + \tau \frac{\pd F}{\pd t}
  - \tau D_1^{} \Delta F,
\end{multline}
which is the same as the equation~\eqref{eq:BoltzEqJeffApprox}.
Initial conditions for this equation are
\begin{equation}
  \label{eq:InitCondBoltzEqJeffAproxSource}
  \left. u \right|_{t=0}^{} = u_0^{},
  \quad
  \left. \frac{\pd u}{\pd t} \right|_{t=0}^{} =
  - \gamma u_0^{}
  - \diverg \bs{J}_0^{}
  + F_0^{},
\end{equation}
where $F_0^{} \equiv F_0^{}(\bs{x}) = \left. F \right|_{t=0}^{}$ is the distribution of sources at time $t=0$.

In the absence of sources and absorption, \ie, if $F = 0$ and $\kappa \equiv \gamma = 0$, the equation \eqref{eq:JeffEqBoltzModelAbsorpSource} takes the form
\begin{equation}
  \label{eq:JeffEqBoltzModelNoAbsorpSource}
  \tau \frac{\pd^2 u}{\pd t^2}
  + \frac{\pd u}{\pd t}
  - \tau D_1^{} \frac{\pd \Delta u}{\pd t}
  - \left( D_1^{} + D_2^{} \right) \Delta u
  =
  0,
\end{equation}
which is the same as the equation~\eqref{eq:JeffEqJeffModelSource} with $f = 0$.

In a steady state the relation \eqref{eq:BoltzEqMomentEqTwoNoAbsorpSourceTauD1D2} takes the form
\begin{equation}
  \label{eq:BoltzEqMomentEqTwoNoAbsorpSourceSteadyTauD1D2}
  \bs{J} =
  - \left( D_1^{} + D_2^{} \right) \grad u
  + \frac{\tau D_1^{}}{4} \left( 3 \Delta \bs{J} + \grad \diverg \bs{J} \right),
\end{equation}
which differs qualitatively from Fick's law.

It is necessary to emphasize that the
relation~\eqref{eq:BoltzEqMomentEqTwoNoAbsorpSourceTauD1D2} is
similar to but not the same as the equation of Guyer and Krumhansl
\eqref{eq:GuKrLawTauD1D2}. The reason is that Guyer and Krumhansl
considered the \emph{linearized} Boltzmann equation rather than the
\emph{linear} one, the difference is explained in
Ref.\,\cite{Cercignani:1988}.

\section{Initial value problems for the homogeneous equation of the Jeffreys type with absorption}
\label{sec:IVPJeffEqNoSource}

In this section we study the classic one-dimensional transport
problem for a substance initially confined in a point. We suppose that
sources are absent, and the substance is absorbed (degraded),
therefore, we set $f = -\gamma u$, where $\gamma$ is the absorption
(degradation) rate. The study reveals a remarkable phenomenon when a
finite portion of the substance \emph{does not move}
though this portion diminishes exponentially with time.

There is also qualitative difference between the two cases $D_2^{} > 0$ and $D_2^{} < 0$. In the first case the solution is wave-like because the characteristic values take complex values. In the second case the characteristic values are real, and, hence, the solution is not wave-like.

\subsection{The Jeffreys type model}

In the one-dimensional case the problem \eqref{eq:JeffEqJeffModelSource}, \eqref{eq:InitCondTeleEqSource} with $f = - \gamma u$ takes the form
\begin{multline}
  \label{eq:JeffEqJeffModelAbsorpOneDim}
  \tau \frac{\pd^2 u_\jeje^{}}{\pd t^2} +
  \left( 1 + \tau \gamma \right) \frac{\pd u_\jeje^{}}{\pd t} -
  \tau D_1^{} \frac{\pd^3 u_\jeje^{}}{\pd x^2 \pd t}\\
  - \left( D_1^{} + D_2^{} \right) \frac{\pd^2 u_\jeje^{}}{\pd x^2}
  + \gamma u_\jeje^{} =
  0,
  \enskip x \in \mbb{R},
  \enskip t>0,
\end{multline}
\begin{equation}
  \label{eq:InitCondJeffEqJeffModelAbsorpOneDim}
  \left. u_\jeje^{} \right|_{t=0}^{} =
  u_0^{},
  \quad
  \left. \frac{\pd u_\jeje^{}}{\pd t} \right|_{t=0}^{} =
  - \gamma u_0^{} - \frac{\pd J_0^{}}{\pd x}.
\end{equation}

The one-dimensional continuity equation \eqref{eq:ContinEqSource} implies that if and only if $\gamma = 0$, \ie, absorption is absent, mass is conserved: $\int_{-\infty}^{\infty} u_\jeje^{}(x,t) \diff x = \int_{-\infty}^{\infty} u_0^{}(x) \diff x \equiv \const$.

Consider the particular initial conditions
\begin{equation}
  \label{eq:InitCondDeltaJeffEqJeffModelAbsorpOneDim}
  u_0^{}(x) = \delta(x)
  \quad\text{and}\quad
  J_0^{}(x) = 0.
\end{equation}

The Fourier transform of the problem \eqref{eq:JeffEqJeffModelAbsorpOneDim}, \eqref{eq:InitCondJeffEqJeffModelAbsorpOneDim}, \eqref{eq:InitCondDeltaJeffEqJeffModelAbsorpOneDim} yields
\begin{multline}
  \label{eq:FourJeffEqJeffModelAbsorpOneDim}
  \tau \frac{\pd^2 \Four u_\jeje^{}}{\pd t^2} +
  \left[ 1 + \tau \left( D_1^{} \xi^2 + \gamma \right) \right] \frac{\pd \Four u_\jeje^{}}{\pd t}\\
  + \left[ \left( D_1^{} + D_2^{} \right) \xi^2 + \gamma \right] \Four u_\jeje^{} =
  0,
  \enskip \xi \in \mbb{R},
  \enskip t>0,
\end{multline}
\begin{equation}
  \label{eq:FourInitCondJeffEqJeffModelAbsorpOneDim}
  \left. \Four u_\jeje^{} \right|_{t=0}^{} =
  1,
  \quad
  \left. \frac{\pd \Four u_\jeje^{}}{\pd t} \right|_{t=0}^{} =
  - \gamma,
\end{equation}
where
\begin{equation*}
  \Four{u}(\xi,\cdot)
  =
  \int_{-\infty}^\infty u(x,\cdot) \e^{\im \xi x} \diff{x}
\end{equation*}
defines the Fourier transform.
The solution to the original problem is, therefore, given by
\begin{equation*}
  u(x,\cdot)
  =
  \frac{1}{2\pi} \int_{-\infty}^\infty \Four u(\xi,\cdot) \e^{-\im x \xi} \diff\xi.
\end{equation*}

The characteristic values of the equation~\eqref{eq:FourJeffEqJeffModelAbsorpOneDim} are
\begin{multline}
  \label{eq:CharValJeffEqJeffModelAbsorp}
  \lambda_{1,2}^{}(\xi)
  =
  \frac{1}{2 \tau} \bigg\{ - \left[ 1 + \tau \left( D_1^{} \xi^2 + \gamma \right) \right]\\
  \pm \sqrt{\left[ 1 - \tau \left( D_1^{} \xi^2 + \gamma \right) \right]^2 - 4 \tau D_2^{} \xi^2} \bigg\},
\end{multline}
where the plus sign corresponds to $\lambda_1^{}$. Note that, if $- D_1^{} < D_2^{} \leq 0$, the characteristic values are real, otherwise, if $D_2^{} >0$, there are two intervals on the real line, symmetric with respect to the origin, where the characteristic values are complex conjugate.

The solution to the problem~\eqref{eq:FourJeffEqJeffModelAbsorpOneDim},~\eqref{eq:FourInitCondJeffEqJeffModelAbsorpOneDim} is
\begin{multline}
  \label{eq:FourSolJeffEqJeffModelAbsorpOneDim}
  \Four u_\jeje^{}(\xi,t)
  =
  \frac{1}{\lambda_1^{}(\xi) - \lambda_2^{}(\xi)}\\
  \times \Big\{ \Big[ \lambda_1^{}(\xi) \e^{\lambda_2^{}(\xi) t} - \lambda_2^{}(\xi) \e^{\lambda_1^{}(\xi) t} \Big]\\
  - \gamma \Big[ \e^{\lambda_1^{}(\xi) t} - \e^{\lambda_2^{}(\xi) t} \Big] \Big\}.
\end{multline}

The asymptotic behaviour of the characteristic values is described by
\begin{subequations}
  \label{eq:AsyCharValJeffEqJeffModelAbsorp}
  \begin{alignat}{2}
    \lambda_1^{}(\xi) & =
    - \frac{1}{\tau} \left( 1 + \frac{D_2^{}}{D_1^{}} \right) + O \!\left( \frac{1}{\xi^2} \right)\notag\\[1ex]
    &\equiv
    -k_2^{} + O \!\left( \frac{1}{\xi^2} \right),\\[1ex]
    \lambda_2^{}(\xi) & =
    - D_1^{} \xi^2 + \frac{D_2^{}}{\tau D_1^{}} - \gamma + O \!\left( \frac{1}{\xi^2} \right)\notag\\[1ex]
    &\equiv
    - \left( D_1^{} \xi^2 + k_1^{} + \gamma \right) + O \!\left( \frac{1}{\xi^2} \right)
  \end{alignat}
\end{subequations}
as $\xi \to \pm\infty$,
where we have used the relations \eqref{eq:RelCoeffJeffTwoPhModels} between the coefficients of the two models.
Therefore, the asymptotic behaviour of the Fourier transform \eqref{eq:FourSolJeffEqJeffModelAbsorpOneDim} with respect to $\xi$ is
\begin{equation}
  \label{eq:AsyFourSolJeffEqJeffModelAbsorpOneDim}
  \Four u_\jeje^{}(\xi,t) =
  \e^{-k_2^{} t} + O \!\left( \frac{1}{\xi^2} \right)
  \quad\text{as}\quad \xi \to \pm\infty.
\end{equation}
This means that the solution $u_\jeje^{}$ has the form
\begin{equation}
  \label{eq:SolIVPJeffEqJeffModelAbsorpOneDim}
  u_\jeje^{}(x,t) =
  u_\jeje^\singul(x,t) + u_\jeje^\regul(x,t),
\end{equation}
where
\begin{equation}
  \label{eq:SolIVPSingJeffEqJeffModelAbsorpOneDim}
  u_\jeje^\singul(x,t) =
  \e^{-k_2^{} t} \delta(x)
\end{equation}
is the singular term, while the regular term $u_\jeje^\regul$ is a continuous function~\cite{Zorich2:2004}. The presence of the singular term means that in the Jeffreys type model a finite portion of the substance \emph{does not move}, though this portion diminishes exponentially with time.

If $\tau \gamma < 1$
the asymptotic behaviour of the regular term with respect to $t$ is
\begin{equation*}
  \e^{\gamma t} \Four u_\jeje^\regul \!\left( \frac{\xi}{\sqrt{t}}, t \right) \to
  \e^{- D_\jeje^{} \xi^2}
  \quad\text{as}\quad t \to +\infty
\end{equation*}
with
\begin{equation*}
  D_\jeje^{} =
  D_1^{} + \frac{D_2^{}}{1 - \tau \gamma},
\end{equation*}
which leads to the asymptotic behaviour
\begin{equation*}
  \e^{\gamma t} \!\sqrt{t} \,u_\jeje^\regul \!\left( \sqrt{t} \hspace{0.1ex}x, t \right) \to
  \frac{1}{2 \sqrt{\pi D_\jeje^{}}} \exp \!\left( \!- \frac{x^2}{4 D_\jeje^{}} \right)
\end{equation*}
as $t \to +\infty$.
This means that if $\tau \gamma < 1$ the solution $u_\jeje^{}$ behaves asymptotically as $t \to +\infty$ as the solution
\begin{equation}
  \label{eq:SolIVPDiffEqD1D2}
  u_\de^{}(x,t)
  =
  \frac{1}{2 \sqrt{\pi D_\jeje^{} t}} \exp \!\left( - \frac{x^2}{4 D_\jeje^{} t} - \gamma t \right)
\end{equation}
of the diffusion equation
\begin{equation*}
  \frac{\pd u_\de^{}}{\pd t} -
  D_\jeje^{} \frac{\pd^2 u_\de^{}}{\pd^2 x} +
  \gamma u_\de^{} =
  0,
  \quad
  x \in \mbb{R},
  \quad
  t>0,
\end{equation*}
with the initial condition
\begin{equation}
  \label{eq:InitCondDiffEqDeltaOneDim}
  \left. u_\de^{} \right|_{t=0}^{} =
  \delta(x).
\end{equation}

\begin{figure*}[!htb]
  \centering
  \includegraphics{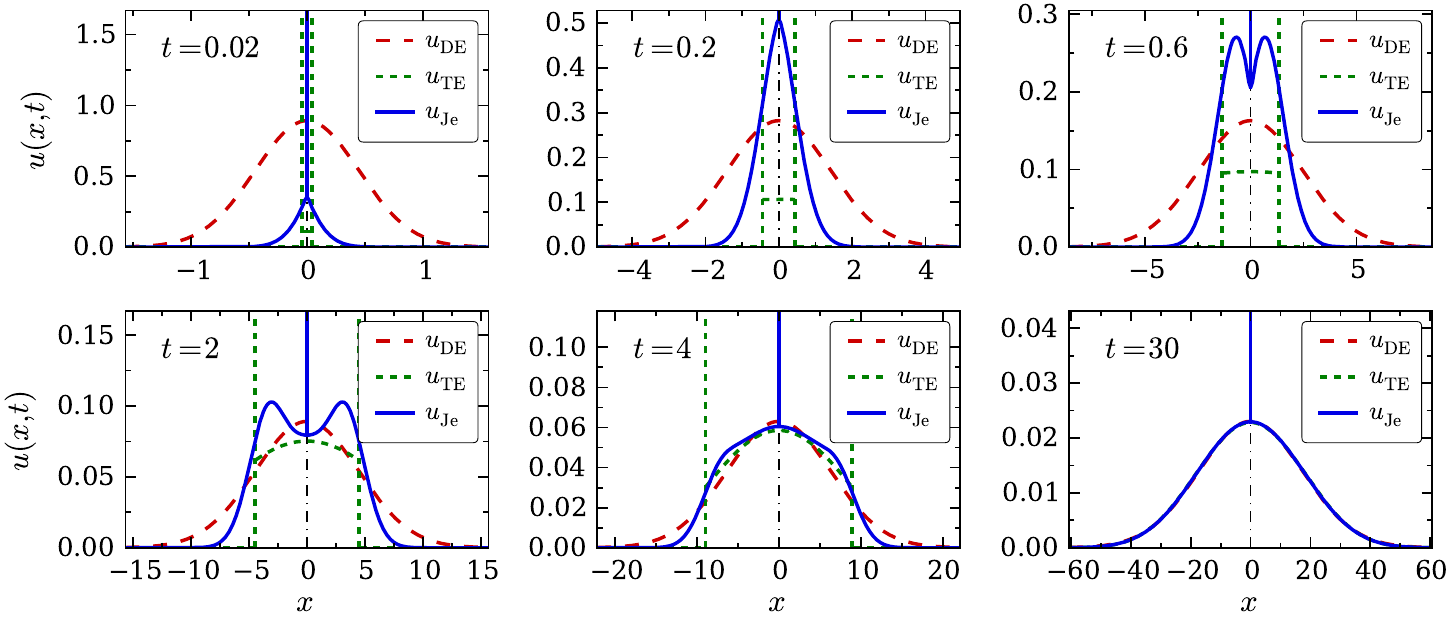}
  \caption{(Color online) The solution $u_\jeje^{}$ to the problem~\eqref{eq:JeffEqJeffModelAbsorpOneDim}, \eqref{eq:InitCondJeffEqJeffModelAbsorpOneDim}, \eqref{eq:InitCondDeltaJeffEqJeffModelAbsorpOneDim} with $\tau = 1$, $D_1^{} = 1$, $D_2^{} = 4$ and $\gamma = 0$ in comparison with those of the diffusion and telegraph equations (see the text). The vertical lines stand for the singular terms $u_\te^\singul$ and $u_\jeje^\singul$. The portions of the regular term are $\int_{-\infty}^{\infty} u_\jeje^\regul(x,0.02) \diff x \approx 0.10$, $\int_{-\infty}^{\infty} u_\jeje^\regul(x,0.2) \diff x \approx 0.63$, $\int_{-\infty}^{\infty} u_\jeje^\regul(x,0.6) \diff x \approx 0.95$ and $\int_{-\infty}^{\infty} u_\jeje^\regul(x,2) \diff x \approx 1.0000$.}
  \label{fig:NoSource_JeffM_D2=4}
\end{figure*}

\begin{figure*}[!htb]
  \centering
  \includegraphics*[width=\linewidth, bb= 0 0 2150 850]{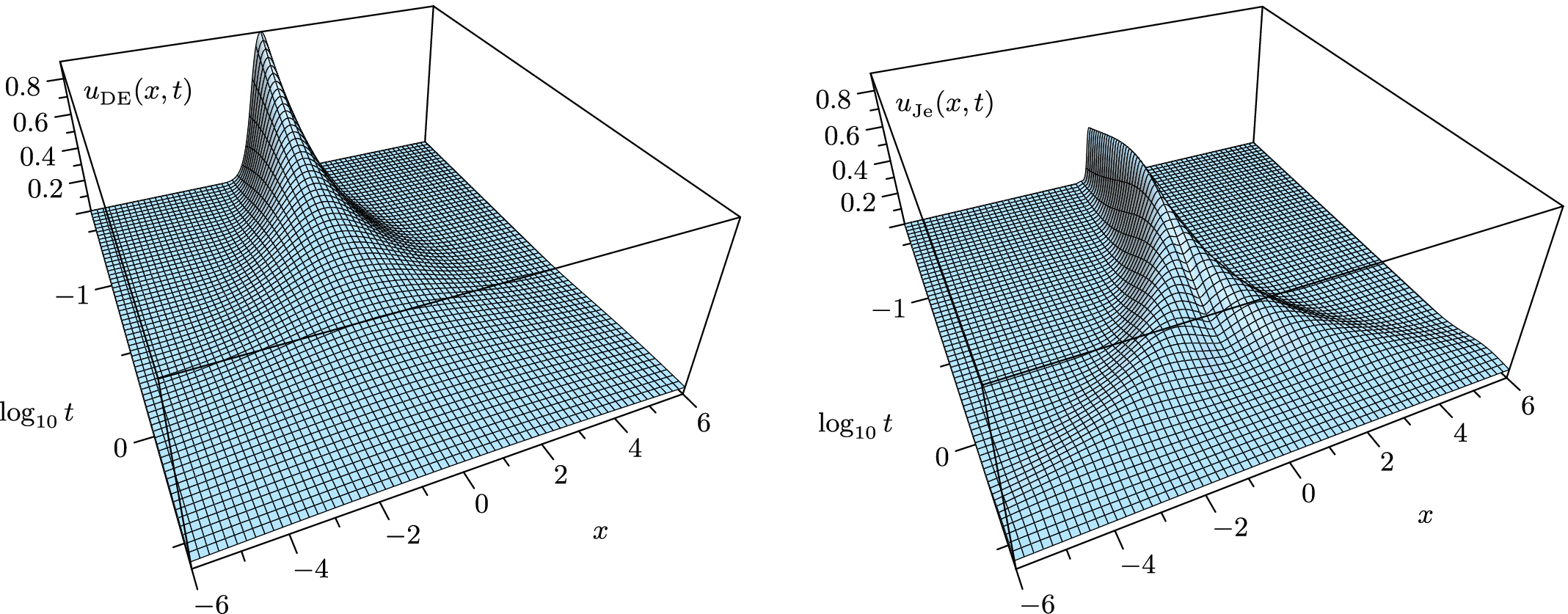}
  \caption{(Color online) The spatio-temporal images of solutions $u_\de^{}$ and $u_\jeje^\regul$ as in Fig.\,\ref{fig:NoSource_JeffM_D2=4}, $t \in [0.02,4]$.}
  \label{fig:No_source_3D_D2=4}
\end{figure*}

\begin{figure}[!htb]
  \centering
  \includegraphics[width=\linewidth]{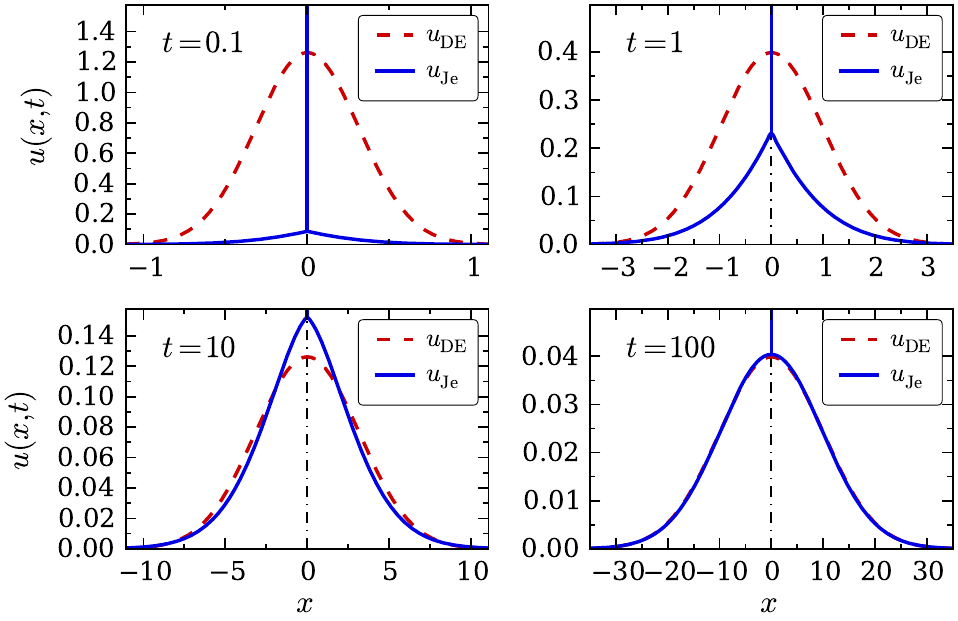}
  \caption{(Color online) The solution $u_\jeje^{}$ to the problem~\eqref{eq:JeffEqJeffModelAbsorpOneDim}, \eqref{eq:InitCondJeffEqJeffModelAbsorpOneDim}, \eqref{eq:InitCondDeltaJeffEqJeffModelAbsorpOneDim} with $\tau = 1$, $D_1^{} = 1$, $D_2^{} = -0.5$ and $\gamma = 0$ in comparison with that of the diffusion equation (see the text). The vertical lines stand for the singular term $u_\jeje^\singul$. The portions of the regular term are $\int_{-\infty}^{\infty} u_\jeje^\regul(x,0.1) \diff x \approx 0.05$, $\int_{-\infty}^{\infty} u_\jeje^\regul(x,1) \diff x \approx 0.39$ and $\int_{-\infty}^{\infty} u_\jeje^\regul(x,10) \diff x \approx 0.99$.}
  \label{fig:NoSource_JeffM_D2=-05}
\end{figure}

Figs.\,\ref{fig:NoSource_JeffM_D2=4}, \ref{fig:No_source_3D_D2=4} and \ref{fig:NoSource_JeffM_D2=-05} show two solutions $u_\jeje^{}$.
In both cases absorption is absent, \ie, $\gamma = 0$, and, therefore, the mass of the substance is conserved.
The solution, shown in Figs.\,\ref{fig:NoSource_JeffM_D2=4} and \ref{fig:No_source_3D_D2=4}, is obtained with the parameters $\tau = 1$, $D_1^{} = 1$ and $D_2^{} = 4$.
The solution, shown in Fig.\,\ref{fig:NoSource_JeffM_D2=-05}, is obtained with the parameters $\tau = 1$, $D_1^{} = 1$ and $D_2^{} = -0.5$ (this corresponds to $k_1^{} = 0.5$ and $k_2^{} = 0.5$).
The solution, shown in Fig.\,\ref{fig:NoSource_JeffM_D2=4}, is wave-like because the characteristic values~\eqref{eq:CharValJeffEqJeffModelAbsorp} take complex values due to $D_2^{} > 0$. In the second case $D_2^{} < 0$, therefore, the characteristic values are real, and for this reason
the solution, shown in Fig.\,\ref{fig:NoSource_JeffM_D2=-05}, is not wave-like.

All the figures show also the diffusion asymptotics \eqref{eq:SolIVPDiffEqD1D2} with $\gamma = 0$, in this case $D_\jeje^{} = D_1^{} + D_2^{}$.

Fig.\,\ref{fig:NoSource_JeffM_D2=4} shows also the solution
of the telegraph equation
\begin{multline*}
  \tau \frac{\pd^2 u_\te^{}}{\pd t^2}
  + \frac{\pd u_\te^{}}{\pd t}
  - \left( D_1^{} + D_2^{} \right) \frac{\pd^2 u_\te^{}}{\pd x^2}
  = 0,\\
  x \in \mbb{R},
  \quad t>0,
\end{multline*}
[this is Eq.\,\eqref{eq:JeffEqJeffModelAbsorpOneDim} with $\pd_{xxt} u_\jeje^{} = 0$ and $\gamma = 0$]
with the initial conditions
\begin{equation*}
  \left. u_\te^{} \right|_{t=0}^{} = \delta(x),
  \quad
  \left. \frac{\pd u_\te^{}}{\pd t} \right|_{t=0}^{} = 0.
\end{equation*}
This solution is given by
\cite{MorseFeshbach:1953v1, MoninYaglom:1971}
\begin{equation*}
  u_\te^{}(x,t)
  = u_\te^\singul(x,t) + u_\te^\regul(x,t),
\end{equation*}
where
\begin{equation*}
  u_\te^\singul(x,t) =
  \frac{\e^{-t/2\tau}}{2 v} \left[ \delta \left( \frac{x}{v} - t \right) + \delta \left( \frac{x}{v} + t \right) \right]
\end{equation*}
is the singular term,
and
\begin{multline*}
  u_\te^\regul(x,t) =
  \frac{\e^{-t/2\tau}}{2 v} \frac{1}{2 \tau}
  \Bigg[ I_0 \!\left( \frac{1}{2 \tau} \sqrt{t^2 - \frac{x^2}{v^2}} \right)\\
  + t \left( \!\sqrt{t^2 - \frac{x^2}{v^2}} \right)^{\hspace{-0.7ex}-1} I_1 \!\left( \frac{1}{2 \tau} \sqrt{t^2 - \frac{x^2}{v^2}} \right) \Bigg] H \!\left( t - \frac{|x|}{v} \right)
\end{multline*}
is the regular term, where $v = \sqrt{(D_1^{} + D_2^{}) / \tau}$ is the velocity, $I_0$ and $I_1$ are the modified Bessel
functions, $H(\cdot)$ is the Heaviside step function. The regular term is discontinuous at $x = \pm v t$.

The solution~\eqref{eq:FourSolJeffEqJeffModelAbsorpOneDim} and the one-dimensional continuity equation~\eqref{eq:ContinEqSource} with $f = 0$ imply that the Fourier transform of flux is
\begin{equation*}
  \Four J(\xi,t) =
  \frac{\im }{\xi} \frac{\lambda_1^{}(\xi) \lambda_2^{}(\xi)}{\lambda_1^{}(\xi) - \lambda_2^{}(\xi)} \left[ \e^{\lambda_1^{}(\xi) t} - \e^{\lambda_2^{}(\xi) t} \right].
\end{equation*}
The asymptotic behaviour of the Fourier transform of flux is described by
\begin{equation*}
  \left( \Four \frac{\pd J}{\pd x} \right) \!(\xi,t) \equiv
  -\im \xi \,\Four J(\xi,t) \to
  k_2^{} \e^{-k_2^{} t}
\end{equation*}
as $\xi \to \pm\infty$,
and means that flux $J(x,t)$ has a finite discontinuity at $x=0$, equal to $k_2^{} \e^{-k_2^{} t}$, which tends to zero as $t \to +\infty$. Note that flux $J(x,t)$ is an odd function with respect to $x$.

\subsection{The two-phase model}

Here we study the behaviour of the net concentration $w_\jetp^{} = u_\jetp^{} + v_\jetp^{}$, where $u_\jetp^{}$ and $v_\jetp^{}$ are the concentrations of the free and bound phases, respectively.
In the one-dimensional case the problems~\eqref{eq:JeffEqTwoPhModelSourceU}, \eqref{eq:InitCondJeffEqTwoPhModelSourceU} and \eqref{eq:JeffEqTwoPhModelSourceV}, \eqref{eq:InitCondJeffEqTwoPhModelSourceV} with $f = -\gamma u$ lead to the problem
\begin{multline}
  \label{eq:JeffEqTwoPhModelAbsorpOneDimW}
  \frac{\pd^2 w_\jetp^{}}{\pd t^2} +
  \left( k_1^{} + k_2^{} + \gamma \right) \frac{\pd w_\jetp^{}}{\pd t} -
  D_1^{} \frac{\pd^3 w_\jetp^{}}{\pd x^2 \pd t}\\
  - k_2^{} D_1^{} \frac{\pd^2 w_\jetp^{}}{\pd x^2} +
  k_2^{} \gamma w_\jetp^{} =
  0,
  \enskip x \in \mbb{R},
  \enskip t>0,
\end{multline}
\begin{equation}
  \label{eq:InitCondJeffEqTwoPhModelAbsorpOneDimW}
  \left. w_\jetp^{} \right|_{t=0}^{} = u_0^{} + v_0^{},
  \quad
  \left. \frac{\pd w_\jetp^{}}{\pd t} \right|_{t=0}^{} =
  D_1^{} \frac{\pd^2 u_0^{}}{\pd x^2}.
\end{equation}

The total mass of the free and bound phases is conserved if and only if $\gamma = 0$, \ie, absorption is absent: $\int_{-\infty}^{\infty} w_\jetp^{}(x,t) \diff x = \int_{-\infty}^{\infty} \left[ u_0^{}(x) + v_0^{}(x) \right] \diff x \equiv \const$. Indeed, Eqs.\,\eqref{eq:TwoPhModelSource} imply that the total mass obeys the equation $\pd_t^{} w_\jetp^{} - D_1^{} \Delta u_\jetp^{} = -\gamma u_\jetp^{}$, or in the Fourier space $\pd_t^{} \Four w_\jetp^{} + \xi^2 D_1^{} \Four u_\jetp^{} = -\gamma \Four u_\jetp^{}$.

Consider the particular initial conditions
\begin{multline}
  \label{eq:InitCondDeltaJeffEqTwoPhModelAbsorpOneDim}
  u_0^{}(x) = \alpha \,\delta(x)
  \quad\text{and}\quad
  v_0^{}(x) = \left( 1 - \alpha \right) \delta(x),\\
  0 \leq \alpha  \leq 1.
\end{multline}

The Fourier transform of the problem~\eqref{eq:JeffEqTwoPhModelAbsorpOneDimW}, \eqref{eq:InitCondJeffEqTwoPhModelAbsorpOneDimW}, \eqref{eq:InitCondDeltaJeffEqTwoPhModelAbsorpOneDim} yields
\begin{multline}
  \label{eq:FourJeffEqTwoPhModelAbsorpOneDimW}
  \frac{\pd^2 \Four w_\jetp^{}}{\pd t^2} +
  \left( k_1^{} + k_2^{} + \gamma + D_1^{} \xi^2 \right) \frac{\pd \Four w_\jetp^{}}{\pd t}\\
  + k_2^{} \left( D_1^{} \xi^2 + \gamma \right) \Four w_\jetp^{} =
  0,
  \quad
  \xi \in \mbb{R},
  \enskip
  t>0,
\end{multline}
\begin{equation}
  \label{eq:FourInitCondJeffEqTwoPhModelAbsorpOneDimW}
  \left. \Four w_\jetp^{} \right|_{t=0}^{}
  = 1,
  \quad
  \left. \frac{\pd \Four w_\jetp^{}}{\pd t} \right|_{t=0}^{}
  =
  - \alpha D_1^{} \xi^2.
\end{equation}
The characteristic values of the equation \eqref{eq:FourJeffEqTwoPhModelAbsorpOneDimW} are
\begin{multline}
  \label{eq:CharValJeffEqTwoPhModelAbsorp}
  \lambda_{1,2}^{}(\xi)
  =
  \frac{1}{2} \bigg[ - \left( k_1^{} + k_2^{} + \gamma + D_1^{} \xi^2 \right)\\
  \pm \sqrt{\left( k_1^{} - k_2^{} + \gamma + D_1^{} \xi^2 \right)^2 + 4 k_1^{} k_2^{}} \bigg],
\end{multline}
where the plus sign corresponds to $\lambda_1^{}$. These values differ from the characteristic values \eqref{eq:CharValJeffEqJeffModelAbsorp} if $\gamma \neq 0$,
however, their asymptotic behaviour is the same, see \eqref{eq:AsyCharValJeffEqJeffModelAbsorp}.

The solution to the problem~\eqref{eq:FourJeffEqTwoPhModelAbsorpOneDimW},~\eqref{eq:FourInitCondJeffEqTwoPhModelAbsorpOneDimW} is
\begin{multline}
  \label{eq:FourSolJeffEqTwoPhModelAbsorpOneDimW}
  \Four w_\jetp^{}(\xi,t)
  =
  \frac{1}{\lambda_1^{}(\xi) - \lambda_2^{}(\xi)}\\
  \times \Big\{ \Big[ \lambda_1^{}(\xi) \e^{\lambda_2^{}(\xi) t} - \lambda_2^{}(\xi) \e^{\lambda_1^{}(\xi) t} \Big]\\
  - \alpha \Big[ \e^{\lambda_1^{}(\xi) t} - \e^{\lambda_2^{}(\xi) t} \Big] D_1^{} \xi^2 \Big\}.
\end{multline}

Taking into account the asymptotic behaviour of the characteristic values, one can conclude that the asymptotic behaviour of the Fourier transform of the solution is
\begin{equation*}
  \label{eq:AsyFourSolJeffEqTwoPhModelAbsorpOneDimW}
  \Four w_\jetp^{}(\xi,t)
  =
  \left( 1 - \alpha \right) \e^{-k_2^{} t}
  + O \!\left( \frac{1}{\xi^2} \right)
  \quad\text{as}\quad \xi \to \pm\infty.
\end{equation*}
cf. with the asymptotics~\eqref{eq:AsyFourSolJeffEqJeffModelAbsorpOneDim}.
This means that the solution $w_\jetp^{}$ has the form
\begin{equation}
  \label{eq:SolIVPJeffEqTwoPhModelAbsorpOneDimW}
  w_\jetp^{}(x,t) =
  w_\jetp^\singul(x,t) + w_\jetp^\regul(x,t),
\end{equation}
where
\begin{equation}
  \label{eq:SolIVPSingJeffEqTwoPhModelAbsorpOneDimW}
  w_\jetp^\singul(x,t) =
  \left( 1 - \alpha \right) \e^{-k_2^{} t} \delta(x)
\end{equation}
is the singular term, while the regular term $w_\jetp^\regul$ is a continuous function~\cite{Zorich2:2004}. The presence of the singular term means that if $\alpha < 1$ in the two-phase model a finite portion of the substance \emph{does not move}, though this portion diminishes exponentially with time.

The asymptotic behaviour of the Fourier transform \eqref{eq:FourSolJeffEqTwoPhModelAbsorpOneDimW} with respect to $t$ is
\begin{equation*}
  \e^{\gamma_\jetp^{} t} \Four w_\jetp^\regul \!\left( \frac{\xi}{\sqrt{t}}, t \right) \to
  \frac{1}{2} \!\left( 1 + \frac{k_{+}^{}}{s} \right)
  \e^{- D_\jetp^{} \xi^2}
\end{equation*}
as $t \to +\infty$,
where
\begin{multline*}
  D_\jetp^{}
  = \frac{1}{2} \!\left( 1 - \frac{k_{-}^{}}{s} \right) \!D_1^{},
  \quad
  \gamma_\jetp^{}
  = \frac{k_{+}^{} - s}{2},\\
  k_\pm^{}
  = k_1^{} + \gamma \pm k_2^{}
  \quad
  \text{and}\quad
  s
  = \sqrt{k_{-}^2 + 4 k_1^{} k_2^{}}.
\end{multline*}
This means that as $t \to +\infty$ the solution $u_\jetp^{}$ behaves asymptotically as
\begin{equation}
  \label{eq:AsyInftySolJeffEqTwoPhModelW}
  w_\jetp^\infty
  \equiv
  \frac{1}{2} \!\left( 1 + \frac{k_{+}^{}}{s} \right)
  u_\de^{}(x,t),
\end{equation}
where
\begin{equation*}
  u_\de^{}(x,t)
  =
  \frac{1}{2 \sqrt{\pi D_\jetp^{} t}} \exp \!\left( - \frac{x^2}{4 D_\jetp^{} t} - \gamma_\jetp^{} t \right)
\end{equation*}
is the solution of the diffusion equation
\begin{multline*}
  \frac{\pd u_\de^{}}{\pd t}
  - D_\jetp^{} \frac{\pd^2 u_\de^{}}{\pd^2 x}
  + \gamma_\jetp^{} u_\de^{}
  = 0,\\
  x \in \mbb{R},
  \quad
  t>0,
\end{multline*}
with the initial condition \eqref{eq:InitCondDiffEqDeltaOneDim}.
If $\gamma = 0$ then $k_{+}^{} = s$ and
\begin{equation*}
  D_\jetp^{}
  =
  \frac{k_2^{}}{k_1^{} + k_2^{}} D_1^{}
  \equiv
  D_1^{} + D_2^{}.
\end{equation*}

\begin{figure}[!htb]
  \centering
  \includegraphics[width=\linewidth]{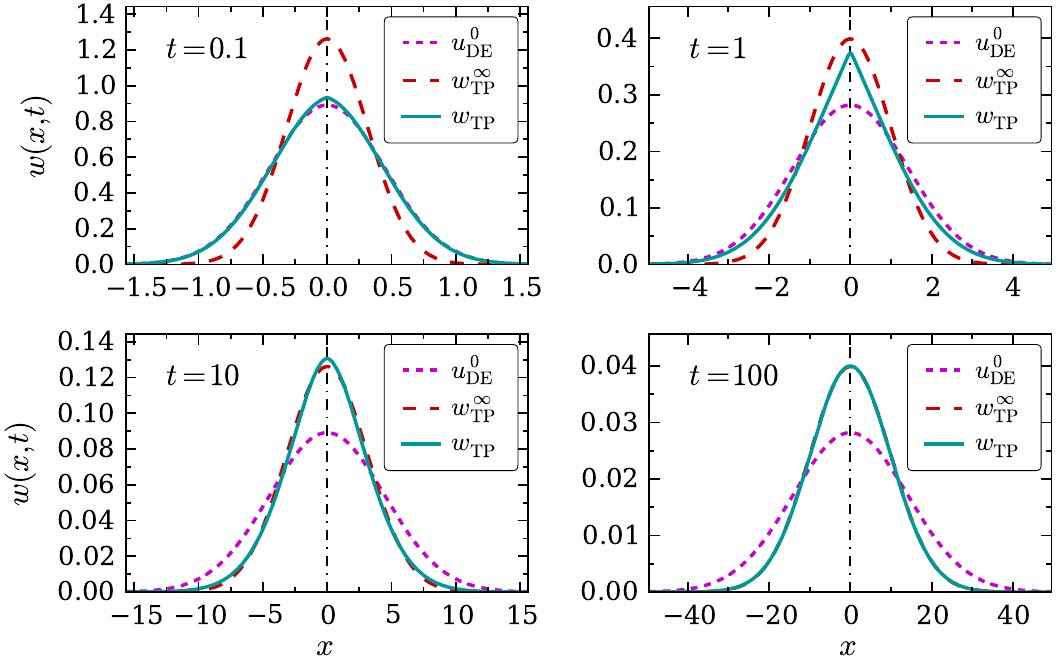}
  \caption{(Color online) The solution $w_\jetp^{}$ to the problem~\eqref{eq:JeffEqTwoPhModelAbsorpOneDimW}, \eqref{eq:InitCondJeffEqTwoPhModelAbsorpOneDimW}, \eqref{eq:InitCondDeltaJeffEqTwoPhModelAbsorpOneDim} with $D_1^{} = 1$, $k_1^{} = 0.5$, $k_2^{} = 0.5$, $\gamma = 0$ and $\alpha = 1$ in comparison with that of the diffusion equation and the diffusion asymptotics $w_\jetp^\infty$ \eqref{eq:AsyInftySolJeffEqTwoPhModelW} with $\gamma = 0$ (see the text).}
  \label{fig:NoSource_TwoPh_D2=-05}
\end{figure}

Fig.\,\ref{fig:NoSource_TwoPh_D2=-05} shows the solution $w_\jetp^{}$ with $\alpha = 1$, \ie, $u_0^{} = \delta(x)$ and $v_0^{} = 0$. The parameters are $D_1^{} = 1$, $k_1^{} = 0.5$, $k_2^{} = 0.5$ (this corresponds to $\tau = 1$ and $D_2^{} = -0.5$, cf. Fig.\,\ref{fig:NoSource_JeffM_D2=-05}) and $\gamma = 0$. The total mass of $w_\jetp^{}$ is conserved. The figure shows also the diffusion asymptotics~\eqref{eq:AsyInftySolJeffEqTwoPhModelW} with $\gamma = 0$, which is the same as the asymptotics~\eqref{eq:SolIVPDiffEqD1D2} with $\gamma = 0$.
For comparison the figure shows the solution $u_\de^{0}$
to the problem for the diffusion equation
\eqref{eq:TwoPhModelSourceEqU} with $k_1^{} = k_2^{} = 0$, $f = -
\gamma u$ and the initial condition $u_0^{} = \delta(x)$, \ie, this would be the concentration of the free substance if interphase mass transfer
were absent.

Note that the solution $w_\jetp^{}$ with $\alpha = 0$, \ie, $u_0^{} = 0$ and $v_0^{} = \delta(x)$, is the same as the solution $u_\jeje^{}$ to the problem~\eqref{eq:JeffEqJeffModelAbsorpOneDim}, \eqref{eq:InitCondJeffEqJeffModelAbsorpOneDim}, \eqref{eq:InitCondDeltaJeffEqJeffModelAbsorpOneDim} with the same parameters, see Fig.\,\ref{fig:NoSource_JeffM_D2=-05}.

\subsection{$D_{N=1}^{}$ approximation to the linear Boltzmann equation}

In the one-dimensional case the equation \eqref{eq:JeffEqBoltzModelAbsorpSource} with $F = 0$ takes the form
\begin{multline*}
  \tau \frac{\pd^2 u_\boltz^{}}{\pd t^2}
  + \left( 1 + \tau \gamma \right) \frac{\pd u_\boltz^{}}{\pd t}
  - \tau D_1^{} \frac{\pd^3 u_\boltz^{}}{\pd x^2 \pd t}\\
  - \left( D_1^{} + D'_2 \right) \frac{\pd^2 u_\boltz^{}}{\pd x^2}
  + \gamma u_\boltz^{}
  = 0,
  \quad
  x \in \mbb{R},
  \quad
  t>0,
\end{multline*}
with $D'_2 = D_2^{} + \tau \gamma D_1^{}$.
The initial conditions \eqref{eq:InitCondBoltzEqJeffAproxSource} in this case are
\begin{equation*}
  \left. u_\boltz^{} \right|_{t=0}^{} =
  u_0^{},
  \quad
  \left. \frac{\pd u_\boltz^{}}{\pd t} \right|_{t=0}^{} =
  - \gamma u_0^{} - \frac{\pd J_0^{}}{\pd x}.
\end{equation*}
This problem is similar to the problem \eqref{eq:JeffEqJeffModelAbsorpOneDim}, \eqref{eq:InitCondJeffEqJeffModelAbsorpOneDim} in the framework of the Jeffreys type model.

\section{The mean square displacement}
\label{sec:MSD}

The mean-square displacement (MSD) is an integral quantity whose temporal dependence characterizes diffusion and Brownian motion.
Of interest is to calculate the MSD in the framework of each model for comparison with that in diffusion and Brownian motion.
This comparison is of particular interest for small $t$ since the asymptotics of the above solutions for large $t$ is difusive.

In this section we calculate the MSD,
defined by $\average{x^2(t)} \equiv \int_{-\infty}^\infty x^2 u(x,t)
\diff x$, for the solutions of the problems, considered in the
previous section, with the initial condition $\left. u
\right|_{t=0}^{} = \delta(x)$. The solution $u(\cdot,t)$ is to be a
probability distribution function for any $t \geq 0$, \ie, the
necessary condition is $\int_{-\infty}^\infty u(x,t) \diff x = 1$.
Therefore, absorption is necessarily absent, \ie, $\gamma = 0$.

Concerning the \emph{diffusion equation it is well
known that} the MSD for the solution to the
problem
\begin{equation*}
  \frac{\pd u_\de^{}}{\pd t} -
  D \frac{\pd^2 u_\de^{}}{\pd^2 x}
  = 0,
  \quad
  x \in \mbb{R},
  \quad
  t>0,
\end{equation*}
\begin{equation*}
  \left. u_\de^{} \right|_{t=0}^{} =
  \delta(x),
\end{equation*}
linearly depends on time and is equal to
\begin{equation*}
  \average{x_\de^2(t)}
  =
  2 D t,
  \quad
  t \geq 0.
\end{equation*}
However, this temporal behaviour of the MSD is wrong at small values of time, where it must be ballistic.

The MSD in the framework of the \emph{Jeffreys type model} is defined through
the solution $u_\jeje^{}$ to the problem~\eqref{eq:JeffEqJeffModelAbsorpOneDim}, \eqref{eq:InitCondJeffEqJeffModelAbsorpOneDim}, \eqref{eq:InitCondDeltaJeffEqJeffModelAbsorpOneDim} with $\gamma = 0$.
Therefore, the MSD is the solution to the problem
\begin{equation*}
  \tau \frac{\diff^2}{\diff t^2} \average{x_\jeje^2}
  + \frac{\diff}{\diff t} \average{x_\jeje^2}
  =
  2 \left( D_1^{} + D_2^{} \right),
  \enskip t>0,
\end{equation*}
\begin{equation*}
  \left. \average{x_\jeje^2} \right|_{t=0} = 0,
  \quad
  \left. \frac{\diff}{\diff t} \average{x_\jeje^2} \right|_{t=0}
  = 0.
\end{equation*}
Hence, the MSD is equal to
\begin{multline}
\label{eq:MSDJeffM}
  \average{x_\jeje^2(t)}
  =
  2 \left( D_1^{} + D_2^{} \right) \left[ t - \tau \left( 1 - \e^{-t/\tau} \right) \right]\\[1.5ex]
  \sim
  \begin{cases}
    \dfrac{D_1^{} + D_2^{}}{\tau} \,t^2 & \text{as}\quad t \to 0,\\[1.5ex]
    2 \left( D_1^{} + D_2^{} \right) t & \text{as}\quad t \to \infty.
  \end{cases}
\end{multline}
The MSD in the framework of the \emph{$D_{N=1}^{}$ approximation to the linear Boltzmann equation} with $\gamma = 0$ is the same.

The temporal behaviour of the MSD given by
Eq.\,\eqref{eq:MSDJeffM} is the same as that in the Brownian motion
described by the standard Langevin equation with initial velocities
having Maxwellian distribution~\cite{Risken:1989, Mazo:2002,
CoffeyEtAl:2004}: it is ballistic as $t \to 0$ and diffusive as $t
\to \infty$.
Therefore, the behaviour of the MSD in the two models is consistent with that in the normal diffusion and Brownian motion.

The MSD in the framework of the \emph{two-phase model} is defined through
the solution $w_\jetp^{}$ to the problem~\eqref{eq:JeffEqTwoPhModelAbsorpOneDimW}, \eqref{eq:InitCondJeffEqTwoPhModelAbsorpOneDimW}, \eqref{eq:InitCondDeltaJeffEqTwoPhModelAbsorpOneDim} with $\gamma = 0$.
Therefore, the MSD is the solution to the problem
\begin{equation*}
  \tau \frac{\diff^2}{\diff t^2} \average{x_\jetp^2}
  + \frac{\diff}{\diff t} \average{x_\jetp^2}
  =
  2 \left( D_1^{} + D_2^{} \right),
  \enskip t>0,
\end{equation*}
\begin{equation*}
  \left. \average{x_\jetp^2} \right|_{t=0} = 0,
  \quad
  \left. \frac{\diff}{\diff t} \average{x_\jetp^2} \right|_{t=0}
  = 2 \alpha D_1^{}.
\end{equation*}
Hence, the MSD is equal to
\begin{multline*}
  \average{x_\jetp^2(t)}
  =
  2 \Big\{ \left( D_1^{} + D_2^{} \right) \Big[ t - \tau \left( 1 - \e^{-t/\tau} \right) \Big]\\
  +
  \alpha D_1^{} \tau \Big( 1 - \e^{-t/\tau} \Big) \Big\},
  \quad
  t \geq 0,
\end{multline*}
where $D_2^{}$ and $\tau$ are given by the relations~\eqref{eq:RelCoeffJeffTwoPhModelsA}. This differs from the behaviour~\eqref{eq:MSDJeffM} if $\alpha > 0$.

\section{Initial value problems for the equation of the Jeffreys type with absorption and stationary point source}
\label{sec:IVPJeffEqSource}

In this section we study the classic problem on the  diffusion of a
substance from a stationary point source. We suppose that the
substance is absorbed (degraded), therefore, we set $f = -\gamma u +
\delta(x)$. We suppose also that the initial concentrations and flux
are equal to zero.

The study reveals that in the model described by the $D_{N=1}^{}$ approximation to the linear Boltzmann equation a finite portion of the substance \emph{does not move},
and this portion \emph{increases} with time, approaching a limit.
Alternatively, in the Jeffreys type and two-phase models the substance \emph{does always move}.

\subsection{The Jeffreys type model}

In the one-dimensional case the problem \eqref{eq:JeffEqJeffModelSource}, \eqref{eq:InitCondTeleEqSource} with $f = - \gamma u + \delta(x)$, $u_0^{} = 0$ and $J_0^{} = 0$ takes the form
\begin{multline}
  \label{eq:JeffEqJeffModelAbsorpDeltaSourceOneDim}
  \tau \frac{\pd^2 u_\jeje^{}}{\pd t^2} +
  \left( 1 + \tau \gamma \right) \frac{\pd u_\jeje^{}}{\pd t} -
  \tau D_1^{} \frac{\pd^3 u_\jeje^{}}{\pd x^2 \,\pd t}\\
  - \left( D_1^{} + D_2^{} \right) \frac{\pd^2 u_\jeje^{}}{\pd x^2} +
  \gamma u_\jeje^{} =
  \delta(x),
  \quad
  x \in \mbb{R},
  \enskip
  t>0,
\end{multline}
\begin{equation}
  \label{eq:InitCondJeffEqJeffModelHomoDeltaSourceOneDim}
  \left. u_\jeje^{} \right|_{t=0}^{} =
  0,
  \quad
  \left. \frac{\pd u_\jeje^{}}{\pd t} \right|_{t=0}^{} =
  \delta(x).
\end{equation}
The Fourier transform of this problem yields
\begin{multline*}
  \tau \frac{\pd^2 \Four u_\jeje^{}}{\pd t^2} +
  \left[ 1 + \tau \left( D_1^{} \xi^2 + \gamma \right) \right] \frac{\pd \Four u_\jeje^{}}{\pd t}\\[1ex]
  + \left[ \left( D_1^{} + D_2^{} \right) \xi^2 + \gamma \right] \Four u_\jeje^{} =
  1,
  \quad
  \xi \in \mbb{R},
  \quad
  t>0,
\end{multline*}
\begin{equation*}
  \left. \Four u_\jeje^{} \right|_{t=0}^{} =
  0,
  \quad
  \left. \frac{\pd \Four u_\jeje^{}}{\pd t} \right|_{t=0}^{} =
  1.
\end{equation*}
The solution to the latter problem is
\begin{multline*}
  \Four u_\jeje^{}(\xi,t) =
  \frac{1}{\lambda_1^{}(\xi) - \lambda_2^{}(\xi)}
  \times\\
  \bigg\{ \frac{1}{\tau} \left[ \frac{\e^{\lambda_1^{}(\xi) t} - 1}{\lambda_1^{}(\xi)} - \frac{\e^{\lambda_2^{}(\xi) t} - 1}{\lambda_2^{}(\xi)} \right] +
    \left[ \e^{\lambda_1^{}(\xi) t} - \e^{\lambda_2^{}(\xi) t} \right] \bigg\},
\end{multline*}
where $\lambda_{1,2}^{}$ are the characteristic values, given by Eq.\,\eqref{eq:CharValJeffEqJeffModelAbsorp}.

The asymptotic behaviour~\eqref{eq:AsyCharValJeffEqJeffModelAbsorp} of the characteristic values leads to the asymptotic behaviour
\begin{equation*}
  \Four u_\jeje^{}(\xi,t)
  =
  O \!\left( \frac{1}{\xi^2} \right)
  \quad\text{as}\quad
  \xi \to \pm\infty.
\end{equation*}
Hence the solution $u_\jeje^{}$ is a
continuous function of $x$~\cite{Zorich2:2004}.

The mass of the substance is equal at any time to that in the similar problems for the diffusion and telegraph equations:
\begin{multline}
  \label{eq:MassJeffM}
  \int_{-\infty}^\infty u_\de^{}(x,t) \diff x
  =
  \int_{-\infty}^\infty u_\te^{}(x,t) \diff x\\
  =
  \int_{-\infty}^\infty u_\jeje^{}(x,t) \diff x
  =
  \frac{1 - \e^{-\gamma t}}{\gamma},
  \quad
  t \geq 0,
\end{multline}
where $u_\de^{}$
is the solution of the diffusion equation
\begin{multline*}
  \frac{\pd u_\de^{}}{\pd t} -
  \left( D_1^{} + D_2^{} \right) \frac{\pd^2 u_\de^{}}{\pd^2 x} +
  \gamma u_\de^{} =
  \delta(x),\\
  x \in \mbb{R},
  \quad
  t>0,
\end{multline*}
with the initial condition
\begin{equation*}
  \left. u_\de^{} \right|_{t=0}^{}
  = 0,
\end{equation*}
and $u_\te^{}$ is the solution of the telegraph equation
\begin{multline*}
  \tau \frac{\pd^2 u_\te^{}}{\pd t^2}
  + \left( 1 + \tau \gamma \right) \frac{\pd u_\te^{}}{\pd t}
  - \left( D_1^{} + D_2^{} \right) \frac{\pd^2 u_\te^{}}{\pd^2 x}\\
  + \gamma u_\te^{} =
  \delta(x),
  \quad
  x \in \mbb{R},
  \quad
  t>0,
\end{multline*}
with the initial conditions
\begin{equation*}
  \left. u_\te^{} \right|_{t=0}^{} =
  0,
  \quad
  \left. \frac{\pd u_\te^{}}{\pd t} \right|_{t=0}^{} =
  \delta(x).
\end{equation*}
Note that the mass does not depend on $\tau$.

\begin{figure*}[!htb]
  \centering
  \includegraphics{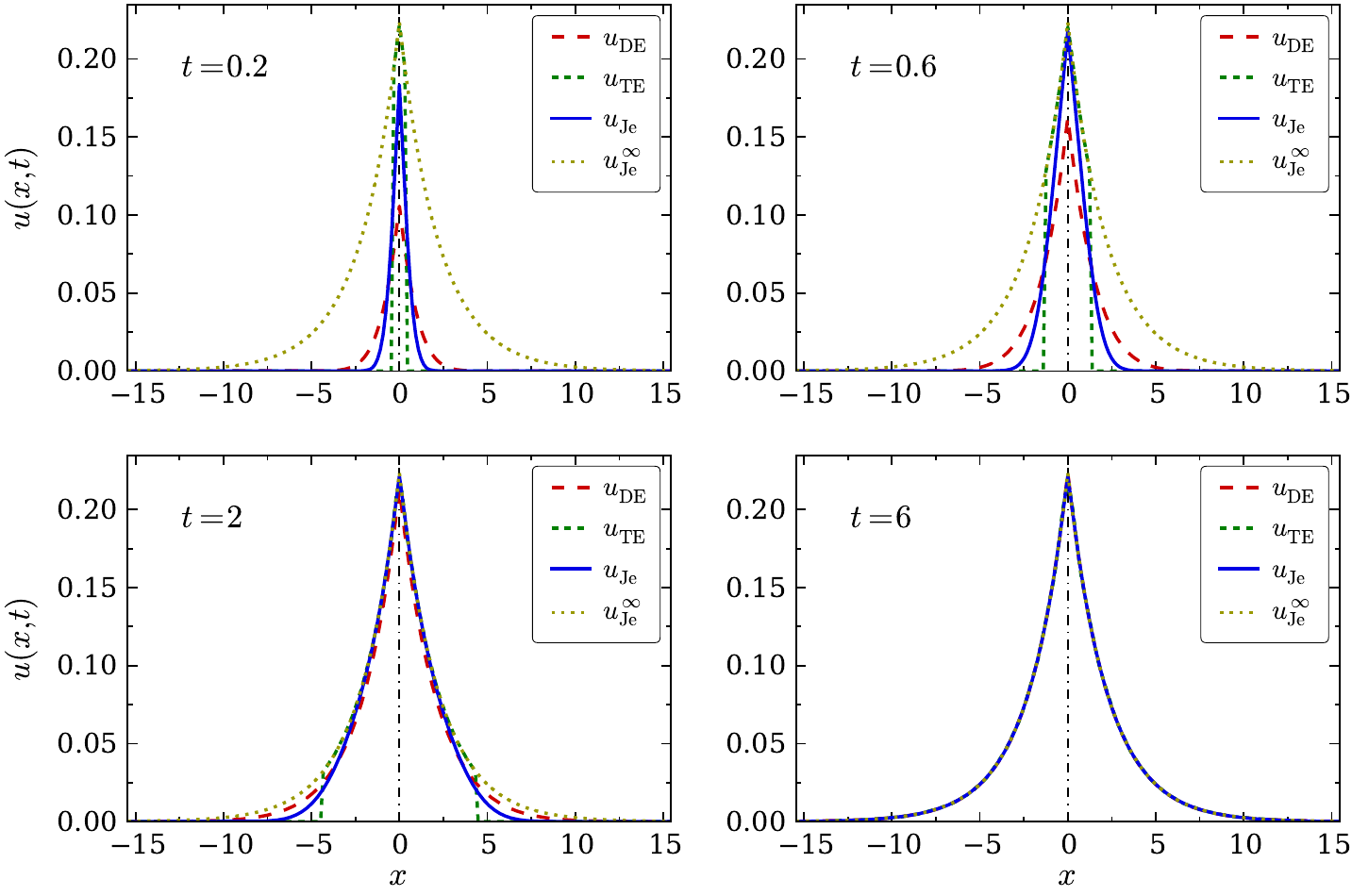}
  \caption{(Color online) The solution $u_\jeje^{}$ to the problem~\eqref{eq:JeffEqJeffModelAbsorpDeltaSourceOneDim}, \eqref{eq:InitCondJeffEqJeffModelHomoDeltaSourceOneDim} with $\tau = 1$, $D_1^{} = 1$, $D_2^{} = 4$ and $\gamma = 1$ in comparison with those of the diffusion and telegraph equations (see the text).}
  \label{fig:Source_JeffM_D2=4}
\end{figure*}

Fig.\,\ref{fig:Source_JeffM_D2=4} shows the solution $u_\jeje^{}$, obtained with the parameters $\tau = 1$, $D_1^{} = 1$, $D_2^{} = 4$ and $\gamma = 1$. The figure shows also the steady state solution $u_\jeje^\steady$ of the equation~\eqref{eq:JeffEqJeffModelAbsorpDeltaSourceOneDim}.

For comparison the figure shows also the solutions $u_\de^{}$ and $u_\te^{}$.
One can see the vertical front
of the solution $u_\te^{}$. The solution $u_\jeje^{}$ is
intermediate between the solutions of the diffusion and telegraph
equations.

\subsection{Two-phase model}

Here we study the behaviour of the net concentration $w_\jetp^{} =
u_\jetp^{} + v_\jetp^{}$, where $u_\jetp^{}$ and $v_\jetp^{}$ are
the concentrations of the free and bound phases, respectively. In
the one-dimensional case the
problems~\eqref{eq:JeffEqTwoPhModelSourceU},
\eqref{eq:InitCondJeffEqTwoPhModelSourceU} and
\eqref{eq:JeffEqTwoPhModelSourceV},
\eqref{eq:InitCondJeffEqTwoPhModelSourceV} with $f = - \gamma u +
\delta(x)$, $u_0^{} = 0$ and $v_0^{} = 0$ lead to the problem
\begin{multline}
  \label{eq:JeffEqTwoPhModelAbsorpDeltaSourceOneDimW}
  \frac{\pd^2 w_\jetp^{}}{\pd t^2} +
  \left( k_1^{} + k_2^{} + \gamma \right) \frac{\pd w_\jetp^{}}{\pd t} -
  D_1^{} \frac{\pd^3 w_\jetp^{}}{\pd x^2 \,\pd t}\\
  - k_2^{} D_1^{} \frac{\pd^2 w_\jetp^{}}{\pd x^2} +
  k_2^{} \gamma w_\jetp^{} =
  \left( k_1^{} + k_2^{} \right) \delta(x),\\
  x \in \mbb{R},
  \quad
  t>0,
\end{multline}
\begin{equation}
  \label{eq:InitCondJeffEqTwoPhModelHomoDeltaSourceOneDimW}
  \left. w_\jetp^{} \right|_{t=0}^{} =
  0,
  \quad
  \left. \frac{\pd w_\jetp^{}}{\pd t} \right|_{t=0}^{} =
  \delta(x).
\end{equation}
Note that the equation~\eqref{eq:JeffEqTwoPhModelAbsorpDeltaSourceOneDimW}, expressed through the parameters $\tau$, $D_1^{}$ and $D_2^{}$, takes the form
\begin{multline*}
  \tau \frac{\pd^2 w_\jetp^{}}{\pd t^2}
  + \left( 1 + \tau \gamma \right) \frac{\pd w_\jetp^{}}{\pd t}
  - \tau D_1^{} \frac{\pd^3 w_\jetp^{}}{\pd x^2 \,\pd t}\\
  - \left( D_1^{} + D_2^{} \right) \frac{\pd^2 w_\jetp^{}}{\pd x^2}
  + \left( 1 + \frac{D_2^{}}{D_1^{}} \right) \gamma w_\jetp^{}
  =
  \delta(x),
\end{multline*}
cf. with the equation~\eqref{eq:JeffEqJeffModelAbsorpDeltaSourceOneDim}, the difference is in the last term of the left-hand side.

The Fourier transform of the problem~\eqref{eq:JeffEqTwoPhModelAbsorpDeltaSourceOneDimW}, \eqref{eq:InitCondJeffEqTwoPhModelHomoDeltaSourceOneDimW}
yields
\begin{multline*}
  \frac{\pd^2 \Four w_\jetp^{}}{\pd t^2} +
  \left( k_1^{} + k_2^{} + \gamma + D_1^{} \xi^2 \right) \frac{\pd \Four w_\jetp^{}}{\pd t}\\
  + k_2^{} \left( D_1^{} \xi^2 + \gamma \right) \Four w_\jetp^{}
  =
  k_1^{} + k_2^{},
  \quad \xi \in \mbb{R},
  \enskip t>0,
\end{multline*}
\begin{equation*}
  \left. \Four w_\jetp^{} \right|_{t=0}^{} =
  0,
  \quad
  \left. \frac{\pd \Four w_\jetp^{}}{\pd t} \right|_{t=0}^{} =
  1.
\end{equation*}
The solution to this problem is
\begin{multline*}
  \Four w_\jetp^{}(\xi,t) =
  \frac{1}{\lambda_1^{}(\xi) - \lambda_2^{}(\xi)}\\
  \times \bigg\{ (k_1^{} + k_2^{}) \bigg[ \frac{\e^{\lambda_1^{}(\xi) t} - 1}{\lambda_1^{}(\xi)} - \frac{\e^{\lambda_2^{}(\xi) t} - 1}{\lambda_2^{}(\xi)} \bigg]\\
  + \left[ \e^{\lambda_1^{}(\xi) t} - \e^{\lambda_2^{}(\xi) t} \right] \bigg\},
\end{multline*}
where $\lambda_{1,2}^{}$ are the characteristic values, given by Eq.\,\eqref{eq:CharValJeffEqTwoPhModelAbsorp}.

The asymptotic behaviour~\eqref{eq:AsyCharValJeffEqJeffModelAbsorp} of the characteristic values leads to the asymptotic behaviour
\begin{equation*}
  \Four w_\jetp^{}(\xi,t)
  =
  O \!\left( \frac{1}{\xi^2} \right)
  \quad\text{as}\quad
  \xi \to \pm\infty.
\end{equation*}
This means that the solution $w_\jetp^{}$ is a continuous function
of $x$~\cite{Zorich2:2004}.

\begin{figure*}[!htb]
  \centering
  \includegraphics{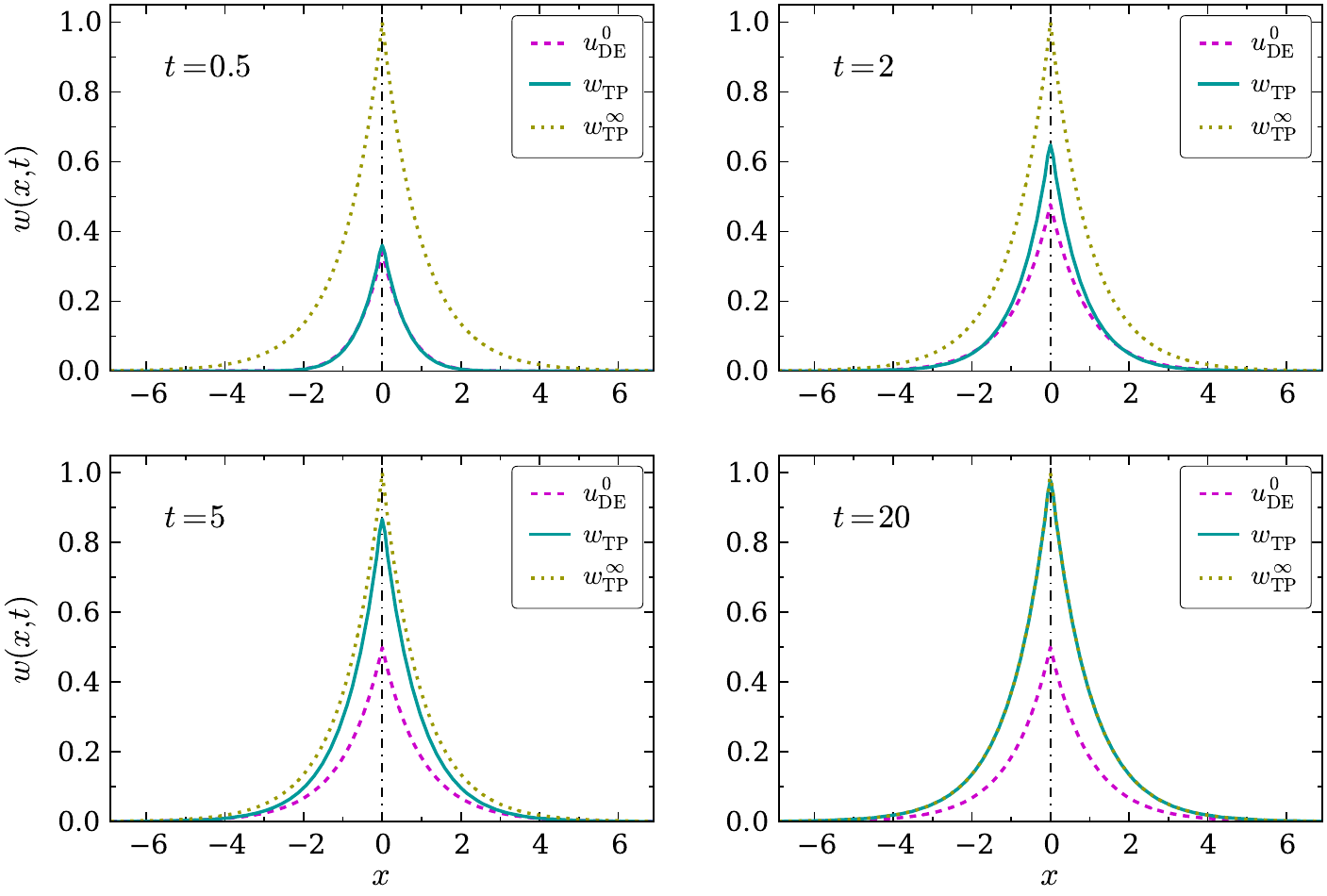}
  \caption{(Color online) The solution $w_\jetp^{}$ to the problem~\eqref{eq:JeffEqTwoPhModelAbsorpDeltaSourceOneDimW}, \eqref{eq:InitCondJeffEqTwoPhModelHomoDeltaSourceOneDimW} with $D_1^{} = 1$, $k_1^{} = 0.5$, $k_2^{} = 0.5$ and $\gamma = 1$ in comparison with that of the diffusion equation (see the text).}
  \label{fig:Source_TwoPh_D2=-05}
\end{figure*}

Fig.\,\ref{fig:Source_TwoPh_D2=-05} shows the solution $w_\jetp^{}$, obtained with the parameters $D_1^{} = 1$, $k_1^{} = 0.5$, $k_2^{} = 0.5$ (this corresponds to $\tau = 1$ and $D_2^{} = -0.5$) and $\gamma = 1$. The figure shows also the steady state solution $w_\jetp^\steady$ of the equation~\eqref{eq:JeffEqTwoPhModelAbsorpDeltaSourceOneDimW}.
For comparison the figure shows the solution $u_\de^{0}$
to the problem for the equation \eqref{eq:TwoPhModelSourceEqU} with $k_1^{} = k_2^{} = 0$, $f = - \gamma u + \delta(x)$ and the homogeneous initial condition, \ie, this would be the concentration of the free substance if the interphase mass transfer were absent.

\subsection{$D_{N=1}^{}$ approximation to the linear Boltzmann equation}

In the one-dimensional case the problem \eqref{eq:JeffEqBoltzModelAbsorpSource}, \eqref{eq:InitCondBoltzEqJeffAproxSource} with $F = \delta(x)$, $u_0^{} = 0$ and $J_0^{} = 0$ takes the form
\begin{multline}
  \label{eq:JeffEqBoltzModelAbsorpDeltaSourceOneDim}
  \tau \frac{\pd^2 u_\boltz^{}}{\pd t^2}
  + \left( 1 + \tau \gamma \right) \frac{\pd u_\boltz^{}}{\pd t}
  - \tau D_1^{} \frac{\pd^3 u_\boltz^{}}{\pd x^2 \pd t}\\
  - \left( D_1^{} + D'_2 \right) \frac{\pd^2 u_\boltz^{}}{\pd x^2}
  + \gamma u_\boltz^{}
  =
  \delta(x)
  - \tau D_1^{} \frac{\pd^2 \delta(x)}{\pd x^2},\\
  \quad
  x \in \mbb{R},
  \quad
  t>0,
\end{multline}
\begin{equation}
  \label{eq:InitCondJeffEqBoltzModelHomoDeltaSourceOneDim}
  \left. u_\boltz^{} \right|_{t=0}^{}
  = 0,
  \quad
  \left. \frac{\pd u_\boltz^{}}{\pd t} \right|_{t=0}^{}
  = \delta(x),
\end{equation}
with $D'_2 = D_2^{} + \tau \gamma D_1^{}$. The Fourier transform of
this problem yields
\begin{multline*}
  \tau \frac{\pd^2 \Four u_\boltz^{}}{\pd t^2}
  + \left[ 1 + \tau \left( D_1^{} \xi^2 + \gamma \right) \right] \frac{\pd \Four u_\boltz^{}}{\pd t}\\
  + \left[ \left( D_1^{} + D'_2 \right) \xi^2 + \gamma \right] \Four u_\boltz^{}
  =
  1 + \tau D_1^{} \xi^2,\\
  \xi \in \mbb{R},
  \quad
  t>0,
\end{multline*}
\begin{equation*}
  \left. \Four u_\boltz^{} \right|_{t=0}^{} =
  0,
  \quad
  \left. \frac{\pd \Four u_\boltz^{}}{\pd t} \right|_{t=0}^{} =
  1,
\end{equation*}
and leads to the following solution:
\begin{multline*}
  \Four u_\boltz^{}(\xi,t) =
  \frac{1}{\lambda_1^{}(\xi) - \lambda_2^{}(\xi)}\\
  \times \bigg\{ \bigg( \frac{1}{\tau} + D_1^{} \xi^2 \bigg) \bigg[ \frac{\e^{\lambda_1^{}(\xi) t} - 1}{\lambda_1^{}(\xi)} - \frac{\e^{\lambda_2^{}(\xi) t} - 1}{\lambda_2^{}(\xi)} \bigg]\\
  + \left[ \e^{\lambda_1^{}(\xi) t} - \e^{\lambda_2^{}(\xi) t} \right] \bigg\},
\end{multline*}
where $\lambda_{1,2}^{}$ are the characteristic values, given by Eq.\,\eqref{eq:CharValJeffEqJeffModelAbsorp} with $D'_2$ instead of $D_2^{}$.

The asymptotic behaviour~\eqref{eq:AsyCharValJeffEqJeffModelAbsorp} of the characteristic values leads to the asymptotic behaviour
\begin{equation*}
  \Four u_\boltz^{}(\xi,t) =
  \frac{1 - \e^{-k'_2 t}}{k'_2} + O \!\left( \frac{1}{\xi^2} \right)
  \quad\text{as}\quad \xi \to \pm\infty,
\end{equation*}
where $k'_2$ is given by the relation~\eqref{eq:RelCoeffJeffTwoPhModelsB} for $k_2^{}$ with $D'_2$ instead of $D_2^{}$, \ie,
\begin{equation*}
  k'_2
  =
  \frac{1}{\tau} \left( 1 + \frac{D'_2}{D_1^{}} \right).
\end{equation*}
Therefore, the solution $u_\boltz^{}$ has the form
\begin{equation*}
  u_\boltz^{}(x,t)
  = u_\boltz^\singul(x,t)
  + u_\boltz^\regul(x,t),
\end{equation*}
where
\begin{equation*}
  u_\boltz^\singul(x,t)
  = \frac{1 - \e^{-k'_2 t}}{k'_2} \delta(x),
\end{equation*}
is the singular term, while the regular term $u_\boltz^\regul$ is a continuous function.
The presence of the singular term means that in this model a finite portion of the substance \emph{does not move},
and this portion \emph{increases} with time up to the value $1 / k'_2$ as $t \to \infty$.

The steady state solution of the equation~\eqref{eq:JeffEqBoltzModelAbsorpDeltaSourceOneDim} satisfies the equation
\begin{equation*}
  - \left( D_1^{} + D'_2 \right) \frac{\pd^2 u_\boltz^\steady}{\pd x^2}
  + \gamma u_\boltz^\steady
  =
  \delta(x)
  - \tau D_1^{} \frac{\pd^2 \delta(x)}{\pd x^2},
\end{equation*}
$x \in \mbb{R}$.
The Fourier transform of the steady state solution is
\begin{multline*}
  \Four u_\boltz^\steady(\xi)
  =
  \frac{1 + \tau D_1^{} \xi^2}{\left( D_1^{} + D'_2 \right) \xi^2 + \gamma}\\
  \equiv \frac{1}{D_1^{} + D_2'} \left[ \tau D_1^{} + \frac{D_1^{} + D_2^{}}{\left( D_1^{} + D'_2 \right) \xi^2 + \gamma} \right].
\end{multline*}
Therefore, the steady state solution is
\begin{equation}
  \label{eq:SolSteadyJeffEqBoltzModelAbsorpDeltaSourceOneDim}
  u_\boltz^\steady(x)
  =
  u_\boltz^\singul(x) + u_\boltz^\regul(x),
\end{equation}
where
\begin{equation}
  \label{eq:SolSteadySingulJeffEqBoltzModelAbsorpDeltaSourceOneDim}
  u_\boltz^\singul(x)
  =
  \frac{\tau D_1^{}}{D_1^{} + D'_2} \delta(x)
  \equiv
  \frac{1}{k'_2} \delta(x),
\end{equation}
is the singular term,
and
\begin{equation}
  \label{eq:SolSteadyRegulJeffEqBoltzModelAbsorpDeltaSourceOneDim}
  u_\boltz^\regul(x)
  =
  \frac{D_1^{} + D_2^{}}{2 \sqrt{\left( D_1^{} + D'_2 \right)^3 \gamma}}
  \exp \!\left( \!- \sqrt{\dfrac{\gamma}{D_1^{} + D'_2}} \left| x \right| \right)
\end{equation}
is the regular (continuous) term.

\begin{figure*}[!htb]
  \centering
  \includegraphics{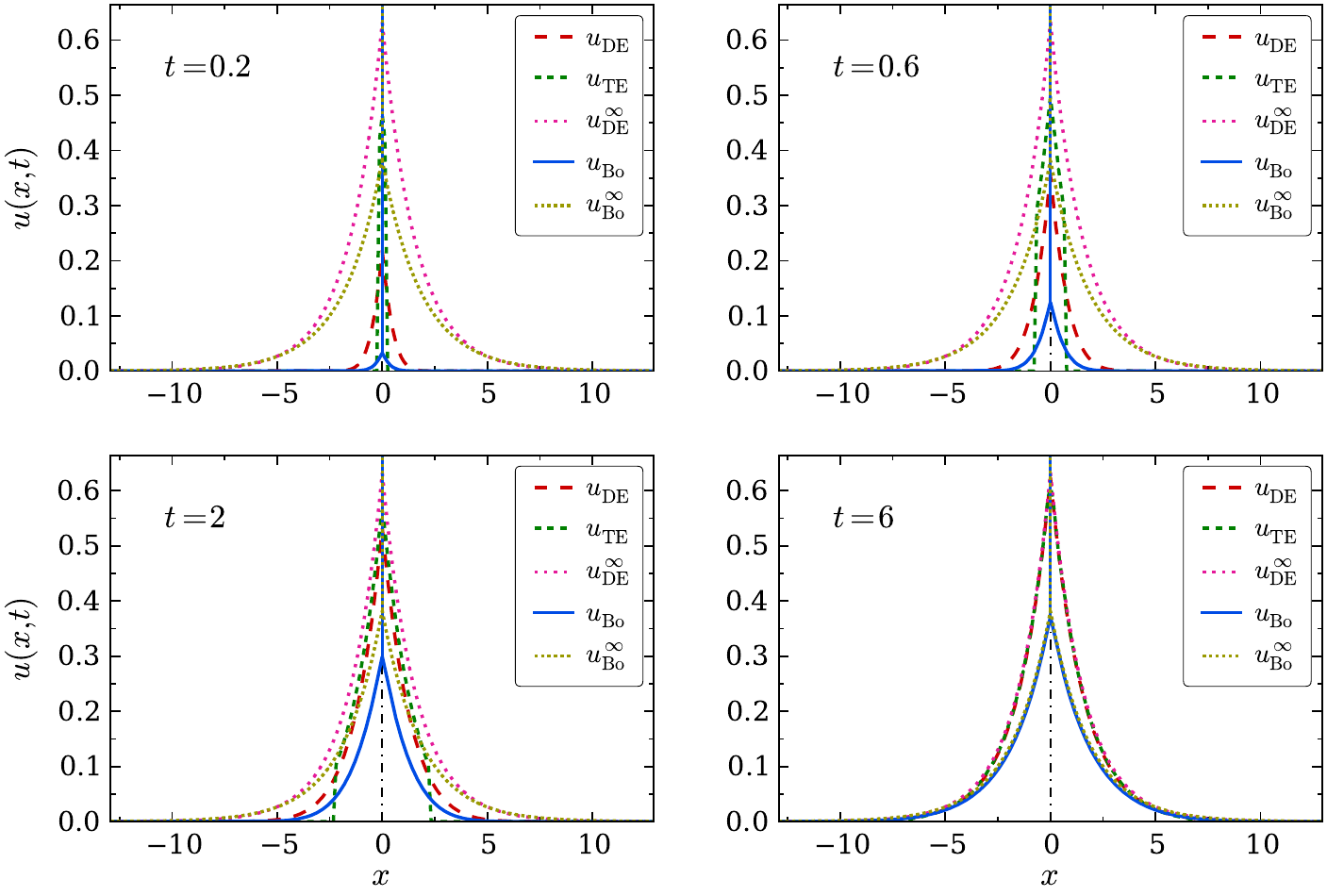}
  \caption{(Color online) The solution $u_\boltz^{}$ to the problem~\eqref{eq:JeffEqBoltzModelAbsorpDeltaSourceOneDim}, \eqref{eq:InitCondJeffEqBoltzModelHomoDeltaSourceOneDim} with $\tau = 1$, $D_1^{} = 1$, $D_2^{} = 0.25$ and $\gamma = 0.5$ in comparison with those of the diffusion and telegraph equations (see the text). The vertical lines stand for the singular term $u_\boltz^\singul$.}
  \label{fig:Source_Boltz_D2=025_gamma=05}
\end{figure*}

Fig.\,\ref{fig:Source_Boltz_D2=025_gamma=05} shows the solution $u_\boltz^{}$, obtained with the parameters $c = \sqrt{15/4}$, $\kappa = 0.5$, $\sigma = 0.5$, isotropic scattering, \ie, $K \equiv (4 \pi)^{-1}$ (therefore, $\sigma_1^{} = \sigma$, $\sigma_2^{} = \sigma$), and $F = \delta(x)$ (this corresponds to $\tau = 1$, $D_1^{} = 1$, $D_2^{} = 0.25$, $D'_2 = 0.75$ and $\gamma = 0.5$). All the figures show also the steady state solution $u_\boltz^\infty$, given by Eqs.\,\eqref{eq:SolSteadyJeffEqBoltzModelAbsorpDeltaSourceOneDim}--\eqref{eq:SolSteadyRegulJeffEqBoltzModelAbsorpDeltaSourceOneDim}.

For comparison the figure shows also the diffusion approximation $u_\de^{}$ to the Boltzmann equation, given by the equation~\eqref{eq:BoltzEqDiffApprox} with the first of the initial conditions~\eqref{eq:InitCondJeffEqBoltzModelHomoDeltaSourceOneDim}, and the ``telegraph'' approximation $u_\te^{}$, given by the equation~\eqref{eq:BoltzEqTeleApprox} with the initial conditions~\eqref{eq:InitCondJeffEqBoltzModelHomoDeltaSourceOneDim}.
The figure shows also the steady state diffusion approximation $u_\de^\infty$.

There are two qualitative peculiarities, which differ
the $D_{N=1}^{}$ approximation from the diffusion and the ``telegraph'' ones. First, a finite portion of the substance in the $D_{N=1}^{}$ approximation \emph{does not move}.
Second, the steady state distribution for the $D_{N=1}^{}$ approximation is different of that for the diffusion and ``telegraph'' approximations.
Nevertheless, the mass of the substance in all these
approximations is the same at any time:
\begin{multline*}
  \int_{-\infty}^\infty u_\de^{}(x,t) \diff x
  =
  \int_{-\infty}^\infty u_\te^{}(x,t) \diff x\\
  =
  \int_{-\infty}^\infty u_\boltz^{}(x,t) \diff x
  =
  \frac{1 - \e^{-\gamma t}}{\gamma},
  \quad
  t \geq 0,
\end{multline*}
cf. with the same relation~\eqref{eq:MassJeffM} in the Jeffreys type model.

\section{Concluding remarks}
\label{sec:Concl}

We have considered three models of non-anomalous mass transfer,
leading to the equation of the Jeffreys type. In the framework of
the Jeffreys type model this equation combines the diffusion and
telegraph ones through the  law of the Jeffreys type, which
combines, in its turn, Fick' law and Cattaneo's equation. In the
framework of the two-phase model the equation of the Jeffreys type
describes the concentrations of the free (mobile) and bound (immobile) phases of a substance as well as the net concentration. The equation of the Jeffreys type in the form of the $D_{N=1}^{}$ approximation ranks
 after the diffusion and telegraph equations in the hierarchy of the spherical harmonics approximations to the linear Boltzmann equation.

Solutions of the equation of the Jeffreys type show qualitatively different behaviour in all these models. The two-phase model shows nothing unusual while the Jeffreys type model and the $D_{N=1}^{}$ approximation to the linear Boltzmann equation exhibit distinctive features.

The first problem we have studied is the transfer
of a substance initially confined in a point. In this case the Jeffreys type model and $D_{N=1}^{}$ approximation coincide. The study has revealed that in these models a finite portion of the substance \emph{does not move},
though this portion diminishes exponentially with time.
Besides, we have calculated the mean square displacement
(MSD) for the solutions of the first problem. The temporal behaviour of the MSD in the Jeffreys type model and in the $D_{N=1}^{}$ approximation is found to be the same as that in the Brownian motion described by the standard Langevin equation, \ie, it is ballistic as $t \to 0$ and diffusive as $t \to \infty$.

The second problem we have studied is the transfer
of a substance from a stationary point source. The study has
revealed that in the $D_{N=1}^{}$ approximation a finite portion of
the substance \emph{does not move}, and this portion
\emph{increases} with time  up to a value as $t \to \infty$.

Tentative interpretation of the local immobilization phenomena is that in a dense crowd inner particles have no space to move, but when the crowd is scattered the particles become mobile.

The two problems we have studied are one-dimensional. An important
question requires further consideration:
are the solutions of the \emph{three-dimensional} problems for the equation of the Jeffreys type left nonnegative?

\section*{Acknowledgements}

We thank the anonymous referee for careful reading of the manuscript and valuable comments, which contributed greatly to the improvement of the paper.

The support by the RFBR grant 11-01-00573-a, the State Contract
no.\,11.519.11.6041 and the EC Collaborative Project
HEALTH-F5-2010-260429 is gratefully acknowledged.

\appendix

\section{Approximations to the linear Boltzmann equation in the framework of the spherical harmonics method}
\label{sec:BoltzEqApprox}

One of the methods to obtain approximate solutions of the linear Boltzmann equation~\eqref{eq:BoltzEq} is the spherical harmonics method \cite{DuderstadtMartin:1979, Modest:2003}. In this method the particle phase space density is expanded into the generalized Fourier series
\begin{equation}
  \label{eq:PhaseSpaceDensSpherHarm}
  \psi(\bs{x},\bs\varOmega,t) =
  \sum_{n=0}^\infty \sum_{m=-n}^n \psi_n^m(\bs{x},t) \,Y_n^m(\bs\varOmega),
\end{equation}
where $Y_n^m$ are the spherical harmonics~\cite{ArfkenWeber:2005},
the coefficients are expressed by
\begin{equation*}
  \psi_n^m(\bs{x},t) =
  \int_\sphere \psi(\bs{x},\bs\varOmega,t) \,\overline{Y_n^m(\bs\varOmega)} \diff\bs\varOmega,
\end{equation*}
where the overline means the complex conjugate.
Note that the expansion~\eqref{eq:PhaseSpaceDensSpherHarm} can be expressed through the particle density $u$ \eqref{eq:BoltzDens} and flux $\bs{J}$ \eqref{eq:BoltzFlux}. Indeed, note that
\begin{equation*}
  \psi_0^0(\bs{x},t) \,Y_0^0 \equiv
  \frac{1}{4 \pi} \,u(\bs{x},t)
\end{equation*}
and
\begin{equation*}
  \sum_{m=-1}^1 \psi_1^m(\bs{x},t) \,Y_1^m(\bs\varOmega) \equiv
  \frac{3}{4 \pi c^2} \,\bs{J}(\bs{x},t) \cdot \bs\varOmega.
\end{equation*}
Therefore, the expansion~\eqref{eq:PhaseSpaceDensSpherHarm} takes the form
\begin{multline*}
  \psi(\bs{x},\bs\varOmega,t) =
  \frac{1}{4 \pi} \,u(\bs{x},t) +
  \frac{3}{4 \pi c^2} \,\bs{J}(\bs{x},t) \cdot \bs\varOmega\\
  + \sum_{n=2}^\infty \sum_{m=-n}^n \psi_n^m(\bs{x},t) \,Y_n^m(\bs\varOmega).
\end{multline*}
The collision kernel is also expanded into the spherical harmonics:
\begin{equation}
  \label{eq:ScatPhaseFuncSpherHarm}
  K(\bs\varOmega \cdot \bs\varOmega') =
  \sum_{n=0}^\infty K_n \sum_{m=-n}^n  Y_n^m(\bs\varOmega) \,\overline{Y_n^m(\bs\varOmega')},
\end{equation}
where
\begin{equation*}
  K_n =
  2 \pi \int_{-1}^1 K(\mu) P_n(\mu) \diff\mu,
\end{equation*}
$P_n$ are the Legendre polynomials, with $K_0 = 1$ due to the normalization $\int_\sphere K(\bs\varOmega \cdot \bs\varOmega') \diff\bs\varOmega = 1$, which is equivalent to $\int_{-1}^1 K(\mu) \diff\mu = (2 \pi)^{-1}$.
The expansions~\eqref{eq:PhaseSpaceDensSpherHarm} and \eqref{eq:ScatPhaseFuncSpherHarm} are substituted into the linear Boltzmann equation. Due to orthogonality of the spherical harmonics this leads to an infinite system of coupled partial differential equations for the functions $\psi_n^m$.

The first equation of the infinite system is the continuity equation \eqref{eq:BoltzEqMomentEqZero}. The second (vector) equation can be obtained with the help of integrating the linear Boltzmann equation, multiplied by $\bs\varOmega$, over the unit sphere. The second equation relates the gradient of the particle density $\grad u$, flux $\bs{J}$, its time derivative $\pd\bs{J} / \pd t$ and coefficients $\psi_2^m$.

\subsection{Diffusion approximation}

The classic \emph{diffusion approximation} is obtained if the coefficients $\psi_n^m$, $n > 1$, in the expansion \eqref{eq:PhaseSpaceDensSpherHarm} are negligible and the coefficients $\psi_1^m$ are quasi-stationary. The latter condition is equivalent to quasi-stationarity of flux $\bs{J}$, \ie, $\pd\bs{J} \!/ \pd t \approx 0$. In this case the particle density and flux are related by the (truncated second) equation
\begin{equation}
  \label{eq:BoltzEqMomentEqOneQuasi}
  \left( \kappa + \sigma_1^{} \right) \bs{J}
  + \frac{c^2}{3} \grad u
  = 0,
\end{equation}
where
\begin{equation*}
  \sigma_n
  = \sigma \left( 1 - K_n \right),
  \quad
  n = 1, 2, \ldots,
\end{equation*}
clearly, $\sigma_n > 0$.
Note that the equation~\eqref{eq:BoltzEqMomentEqOneQuasi} is identical to Fick's law \eqref{eq:FickLaw}.
The continuity equation \eqref{eq:BoltzEqMomentEqZero} and equation \eqref{eq:BoltzEqMomentEqOneQuasi} imply that the particle density satisfies the \emph{diffusion equation}
\begin{equation}
  \label{eq:BoltzEqDiffApprox}
  \frac{\pd u}{\pd t} - \frac{c^2}{3 \left( \kappa + \sigma_1^{} \right)} \Delta u + \kappa u =
  F.
\end{equation}

\subsection{$P_N^{}$ approximations}

The classic $P_N^{}$ approximations are obtained if the coefficients $\psi_n^m$, $n > N \geq 1$, in the expansion \eqref{eq:PhaseSpaceDensSpherHarm} are negligible. Particularly, in the \emph{$P_1$ approximation} the particle density and flux are related by the (truncated second) equation
\begin{equation}
  \label{eq:BoltzEqMomentEqOne}
  \frac{\pd \bs{J}}{\pd t} +
  \left( \kappa + \sigma_1^{} \right) \bs{J} +
  \frac{c^2}{3} \grad u =
  0,
\end{equation}
which is the extension of Eq.\,\eqref{eq:BoltzEqMomentEqOneQuasi}. Note that Eq.\,\eqref{eq:BoltzEqMomentEqOne} is similar
to Cattaneo's equation \eqref{eq:CattEq}. The continuity equation \eqref{eq:BoltzEqMomentEqZero} and equation \eqref{eq:BoltzEqMomentEqOne} imply that the particle density satisfies the \emph{telegraph equation}
\begin{multline}
  \label{eq:BoltzEqTeleApprox}
  \frac{\pd^2 u}{\pd t^2} +
  \left( 2 \kappa + \sigma_1^{} \right) \frac{\pd u}{\pd t} -
  \frac{c^2}{3} \Delta u +
  \kappa \left( \kappa + \sigma_1^{} \right) u\\
  =
  \left( \kappa + \sigma_1^{} \right) F + \frac{\pd F}{\pd t}.
\end{multline}

\subsection{$D_N^{}$ approximations}
\label{sec:BoltzEqD1Approx}

Recently, $D_N^{}$ approximations were proposed~\cite{SchaferEtAl:2011}. They generalize the diffusion approximation, which can be considered as the $D_0^{}$ approximation. The $D_N^{}$ approximations are obtained if the coefficients $\psi_n^m$, $n > N + 1$, in the expansion \eqref{eq:PhaseSpaceDensSpherHarm} are negligible and the coefficients $\psi_{N+1}^m$ are quasi-stationary. The coefficients $\psi_{N+1}^m$ can be expressed through $\psi_N^m$, and the $D_N^{}$ approximation is described by $\psi_n^m$, $n = 0,\ldots,N$. In the case $N=1$ the coefficients $\psi_2^m$ can be expressed through flux $\bs{J}$. As a result, in the \emph{$D_{N=1}^{}$ approximation} the particle density and flux are related by the equation
\begin{multline}
  \label{eq:BoltzEqMomentEqTwo}
  \frac{\pd \bs{J}}{\pd t} +
  \left( \kappa + \sigma_1^{} \right) \bs{J} +
  \frac{c^2}{3} \grad u\\
  =
  \frac{c^2}{15 \left( \kappa + \sigma_2^{} \right)} \left( 3 \Delta \bs{J} + \grad \diverg \bs{J} \right),
\end{multline}
which is the generalization of Eq.\,\eqref{eq:BoltzEqMomentEqOne}.
(We used the notation $D_{N=1}^{}$ instead of $D_1^{}$, since the latter stands for the coefficient.)
The continuity equation \eqref{eq:BoltzEqMomentEqZero} and \eqref{eq:BoltzEqMomentEqTwo} imply that the particle density satisfies the \emph{equation of the Jeffreys type}
\begin{multline}
  \label{eq:BoltzEqJeffApprox}
  \frac{\pd^2 u}{\pd t^2}
  + \left( 2 \kappa + \sigma_1^{} \right) \frac{\pd u}{\pd t}
  - \frac{4c^2}{15 \left( \kappa + \sigma_2^{} \right)} \frac{\pd \Delta u}{\pd t}\\
  - \left[ \frac{c^2}{3} + \frac{4 c^2}{15 \left( \kappa + \sigma_2^{} \right)} \kappa \right] \Delta u
  + \kappa \left( \kappa + \sigma_1^{} \right) u\\
  = \left( \kappa + \sigma_1^{} \right) F
  + \frac{\pd F}{\pd t}
  - \frac{4 c^2}{15 \left( \kappa + \sigma_2^{} \right)} \Delta F.
\end{multline}

\section{Model of Guyer and Krumhansl}

In this section we consider heat transfer. The energy equation
without sources and  sinks has the form
\begin{equation}
  \label{eq:EnergyEq}
  C \frac{\pd T}{\pd t} + \diverg \bs{q} = 0,
\end{equation}
where $T \equiv T(\bs{x},t)$ is temperature, $\bs{q} \equiv \bs{q}(\bs{x},t)$ is heat flux,
$C$ is the volumetric heat capacity.

The equation of Guyer and Krumhansl relating  heat flux and
temperature, is \cite{GuyerKrumhansl:1966a,JosephPreziosi:1989,
Straughan:2011}
\begin{equation}
  \label{eq:GuKrLaw}
  \frac{\pd \bs{q}}{\pd t} + \frac{1}{\tau_R^{}} \bs{q} =
  - \frac{c^2 C}{3} \grad T  + \frac{\tau_N^{} c^2}{5} \left( \Delta \bs{q} + 2 \grad \diverg \bs{q} \right),
\end{equation}
where $c$ is the average speed of phonons, $\tau_N^{}$ is a
relaxation time  for momentum-conserving collisions (normal process)
and $\tau_R^{}$ is a relaxation time for momentum-nonconserving
collisions (``umklapp'' process) in the phonon gas. An equivalent
equation was obtained in the framework of extended irreversible
thermodynamics \cite{JouEtAl:2010}.

The energy equation~\eqref{eq:EnergyEq} and the equation of Guyer and Krumhansl imply that temperature satisfies the homogeneous \emph{equation of the Jeffreys type}
\begin{equation}
  \label{eq:JeffEqGuKrModel}
  \frac{\pd^2 T}{\pd t^2} + \frac{1}{\tau_R^{}} \frac{\pd T}{\pd t} - \frac{3 \tau_N^{} c^2}{5} \frac{\pd \Delta T}{\pd t} - \frac{c^2}{3} \Delta T
  = 0.
\end{equation}
This equation is related to the equation~\eqref{eq:JeffEqBoltzModelNoAbsorpSource} by $\tau = \tau_R^{}$,
$D_1^{} = 3\tau_N^{} c^2/5$ and $D_2^{} = (\tau_R^{}/3 - 3 \tau_N^{}/5) c^2$, besides, the inequalities $D_2^{} \gtrless 0$ are equivalent to $5\tau_R^{} \gtrless 9\tau_N^{}$.
Initial conditions for the equation~\eqref{eq:JeffEqGuKrModel} are
\begin{equation*}
  \left. T \right|_{t=0}^{} = T_0^{},
  \quad
  \left. \frac{\pd T}{\pd t} \right|_{t=0}^{} =
  - \frac{1}{C} \diverg \bs{q}_0^{},
\end{equation*}
where $T_0^{} \equiv T_0^{}(\bs{x})$ and $\bs{q}_0^{} \equiv \bs{q}_0^{}(\bs{x})$ are initial temperature and heat flux, respectively.
These are the same as the initial conditions~\eqref{eq:InitCondBoltzEqJeffAproxSource} with $F = 0$ and $\gamma = 0$.

The equation of Guyer and Krumhansl \eqref{eq:GuKrLaw}, written through $\tau$, $D_1^{}$ and $D_2^{}$, has the form
\begin{multline}
  \label{eq:GuKrLawTauD1D2}
  \tau \frac{\pd \bs{q}}{\pd t} + \bs{q}\\
  =
  - \left( D_1^{} + D_2^{} \right) C \grad T
  + \frac{\tau D_1^{}}{3} \left( \Delta \bs{q} + 2 \grad \diverg \bs{q} \right),
\end{multline}
which differs of the similar equation~\eqref{eq:BoltzEqMomentEqTwoNoAbsorpSourceTauD1D2} in the framework of the $D_{N=1}^{}$ approximation to the linear Boltzmann equation.
In a steady state the equation~\eqref{eq:GuKrLawTauD1D2} takes the form
\begin{equation*}
  \bs{q}
  =
  - \left( D_1^{} + D_2^{} \right) C \grad T
  + \frac{\tau D_1^{}}{3} \left( \Delta \bs{q} + 2 \grad \diverg \bs{q} \right),
\end{equation*}
which differs qualitatively from Fourier's law.

\providecommand{\noopsort}[1]{}\providecommand{\singleletter}[1]{#1}%

\end{document}